\documentclass[12pt,reqno]{amsart}
\allowdisplaybreaks[4]
\usepackage{graphicx} 
\usepackage{epsfig}
\usepackage{amssymb}
\usepackage{epsfig}
\usepackage{amsmath,amsfonts}
\usepackage{cite}

\begin{document}
\title[modified KdV]{Few-cycle optical rogue waves:
complex modified Korteweg-de Vries equation}
\author{Jingsong He\dag$^*$, Lihong Wang\dag, Linjing Li\dag, K.Porsezian\ddag}
\author{R.~Erd\'elyi$\S$}
\thanks{$^*$Corresponding Author: Email:
 hejingsong@nbu.edu.cn, Tel: 86-574-87600739, Fax: 86-574-87600744}
\dedicatory { \dag \ Department of Mathematics, Ningbo University,
Ningbo, Zhejiang 315211, P.\ R.\ China \\
 \ddag Department of Physics, Pondicherry
University, Pondicherry-605014, India\\
\S Solar Physics and Space Plasma Research Centre, University of
Sheffield, Sheffield, S3 7RH, UK}
\begin{abstract}
In this paper, we consider the complex modified
Korteweg-de Vries (mKdV) equation as a model of few-cycle
optical pulses. Using the Lax pair, we construct a generalized
Darboux transformation and systematically generate the first-, second-
and third-order rogue wave solutions and analyze the nature of evolution
of higher-order rogue waves in detail. Based on detailed numerical and
analytical investigations, we classify the higher-order rogue waves
with respect to their intrinsic structure, namely, fundamental
pattern, triangular pattern, and ring pattern. We also present
several new patterns of the rogue wave according to the standard and
non-standard decomposition. The results of this paper explain the generalization of higher-order rogue waves in terms of rational solutions.
 We apply the contour line method to obtain the analytical formulas of the length and width of the
first-order RW of the complex mKdV and the NLS equations.  In nonlinear optics, the higher-order rogue wave solutions found here will be very
   useful to generate high-power few-cycle optical pulses which
   will be applicable in the area of ultra-short pulse technology.
\end{abstract}
 \maketitle

\noindent {{\bf PACS number(s)}: 05.45.Yv, 42.65.Tg, 03.75.Lm,
87.14.gk

\section{Introduction}

The theory of nonlinear dynamics has attracted considerable interest and
is fundamentally linked to several basic developments in the area of
soliton theory. It is well-known that the Korteweg-de Vries (KdV) equation,
modified Korteweg-de Vries (mKdV) equation, sine Gordon equation and
the nonlinear Schr\"odinger (NLS) equation are the most typical and
well-studied integrable evolution equations which describe
nonlinear wave phenomena for a range of dispersive physical
systems. Their stable multi-soliton solutions play an important
role in the study of nonlinear waves\cite{Agra}. Further studies
have also been carried out to examine the effects on these solitons due to dissipation,
inhomogeneity or non-uniformity present in nonlinear media \cite{Malomed,Turitsyn}.

The term ``soliton'' is a sophisticated mathematical concept that
derives its name from the word ``solitary wave'' which is a localized
wave of translation that arises from the balance between nonlinear
and dispersive effects\cite{Agra}. In spite of the initial
theoretical investigations, the concept of solitary wave could not gain
wide recognition for a number of years in the midst of excitement created by
the development of electromagnetic
concepts in those times.  Korteweg and de Vries (1895)
developed a  mathematical model for the shallow water problem and
demonstrated the possibility of solitary wave
generation\cite{Korteweg}. Next, the study of solitary waves really
took off in the mid-1960s when Zabusky and Kruskal discovered the
remarkably stable particle-like behaviour of solitary
waves\cite{Zabusky}. They reported numerical experiments where
solitary waves, described by the KdV equation,  passed through each other
unchanged in speed or shape, which led them to coin the
word ``soliton" to suggest such a unique property. In a follow-up study
Zakharov and Shabat generalized the inverse scattering method in 1972 and
also solved the nonlinear Schr\"odinger equation, demonstrating both its
integrability and the existence of soliton solutions\cite{SPJETP}.

Following the above discoveries, solitary waves of all flavors advanced
rapidly in many areas of science and technology. In nonlinear
physics applications to  many areas e.g. hydrodynamics,
biophysics, atomic physics, nonlinear optics, etc., have been developed.
 As of now, more
than a few hundreds of nonlinear evolution equations (NEEs) have been
shown to admit solitons and some of these theoretical equations are
also responsible for the experimental discovery of
solitons\cite{Agra,Molla}. In general, nonlinear phenomena are often
modelled by nonlinear evolution equations exhibiting a wide range of
high complexities in terms of difference in linear and nonlinear
effects. In the past four decades or so, the advent of high-speed
computers, many advanced mathematical softwares and the development of
a number of sophisticated and systematic analytical methods,  which are
well-supported by experiments have encouraged both  theoreticians and
experimentalists. Nonlinear science has
experienced an explosive growth by the invention of several exciting
and fascinating new concepts not just like solitons, but e.g. dispersion-managed
solitons, rogue waves, similaritons, supercontinuum
generation, etc.\cite{Agra}. Many of the completely integrable
nonlinear partial differential equations (NPDEs) admit one of the
most striking aspects of nonlinear phenomena, which describe soliton
as a universal character and they are of great mathematical as well
as physical interest. It is impossible to discuss all these
manifestations exhaustively in this paper. We further restrict ourselves
to the solitary wave manifestation in nonlinear optics. In the area of
soliton research at the forefront, right now,  is the study of optical
solitons, where the highly sought-after goal is to use strong
localized nonlinear optical pulses as the high-speed
information-carrying bits in optical fibers.

Optical solitons are localized electromagnetic waves that propagate
steadily in a nonlinear medium resulting from the robust balance
between nonlinearity and linear broadening due to dispersion and
diffraction. Existence of the optical soliton was first time found in
1973 when Hasegawa and  Tappert demonstrated the propagation of a
pulse through a nonlinear optical fiber described by the nonlinear
Schr\"odinger equation\cite{Hase1}. They performed a number of
computer simulations demonstrating that nonlinear pulse transmission
in optical fibers would be stable. Subsequently, after the
fabrication of low-loss fiber, Mollenauer et al. in 1980
successfully confirmed this theoretical prediction of soliton
propagation in a laboratory experiment\cite{Molla}. Since then, fiber solitons have
emerged as a very promising potential candidate in long-haul fiber
optic communication systems.

Further, in addition to several important developments in soliton
theory, the concept of modulational instability (MI) has also been
widely used in many nonlinear systems to explain why experiments
involving white coherent light supercontinuum generation (SCG), admit
a triangular spectrum which can be described by the analytical
expressions for the spectra of Akhmediev breather solutions at the
point of extreme compression\cite{Agra}. In the case of the NLS
equation, Peregrine already in \cite{peregrine} had identified the
role of MI in the formation of patterns resembling high-amplitude
freak waves or rogue wave (RW). RWs have recently been also reported in
different areas of science. In particular, in photonic crystal fibre
RWs are well-established in connection with SCG \cite{soli1}.
This actually has stimulated research for RWs in other physical
systems and has paved the way for a number important applications,
including the control of RWs by means of SCG\cite{soli2,dudley}, as
well as studies in e.g. superfluid Helium \cite{ganshin}, Bose
Einstein condensates \cite{konotop1},  plasmas \cite{ruderman,
shukla},  microwave \cite{hohmann},  capillary phenomena
\cite{shats}, telecommunication data streams \cite{vergeles},
inhomogeneous media \cite{arecchi},  water experiments
\cite{akhmediev3}, and so on. Recently, Kibler \emph{et al.}
\cite{akhmediev2} using a suitable experiment with optical fibres
were able to generate femtosecond pulses with strong temporal and
spatial localization and near-ideal temporal Peregrine soliton
characteristics.

For the past couple of years, several nonlinear evolution equations
were shown to exhibit the RW-type rational
solutions
\cite{akhmediev1,akhmediev5,Philippe,akhmediev6,Bandelow1,matveev2,jianke,he1,lu1,guo1,
fabio1,qin1,zhao1,taki1,Shihua,Zhaqilao,liwuwanghe2013}. From the
above listed works, it is clear that one of the possible
generating mechanisms \cite{hezhangwangpf2012} for the higher-order
RW is the interaction of multiple breathers possessing identical and
very particular frequency of the underlying equation. Though the
theory of solitons and many mathematical methods have been well-used
in connection with soliton theory for the past four decades or so,
to the best of our knowledge, the dynamics of multi-rogue wave
evolutions has not yet been systematically investigated in integrable
nonlinear systems \cite{Ablowitz}.

Very recently, considering the propagation of few-cycle optical
pulses in cubic nonlinear media and by developing multiple scaling
approach to the Maxwell-Bloch-Heisenberg equation up to the third-order in
terms of expansion parameter, the complex mKdV equation was derived\cite {leblond,leblond1}. Circularly polarized few-cycle
optical solitons were found which are valid for
long pulses. Thus, it is more than worthy to systematically investigate
the existence of the few-cycle optical rogue waves for this model, and
this is the main purpose of the present paper.

The organization of this paper is as follows. In Section 2, based on
the parameterized Darboux transformation (DT) of the mKdV equation, the
general formation of the solution is given. In Section 3, we
construct the higher-order rogue waves from a periodic seed with
constant amplitude and analyze their structures in detail by
choosing suitable system parameters. We provide detailed
discussion about the obtained results in Sections 4 and 5.

\section{The Darboux Transformation}
 For our analysis, we begin with coupled complex mKdV equations of the form of
\begin{eqnarray}\label{coupledmkdv}
q_t+q_{xxx}-6qrq_x=0,\label{coupledmkdvq}\\
r_t+r_{xxx}-6rqr_x=0. \label{coupledmkdvr}
\end{eqnarray}
Under a reduction condition $q=-r^*$, the above coupled equations
reduce to the complex mKdV
\begin{equation}\label{mkdv}
q_t+q_{xxx}+6|q|^2q_x=0.
\end{equation}
The complex mKdV equation is one of the well-known and completely
integrable equations in soliton theory, which possesses all the
basic characters of integrable models. From a physical point of
view, the above equation has been derived for, e.g. the dynamical
evolution of nonlinear lattices, plasma physics, fluid dynamics,
ultra-short pulses in nonlinear optics, nonlinear transmission lines
and so on\cite{Ablowitz}. The Lax pair corresponding to the coupled mKdV equations
is given by\cite{Ablowitz}, i.e.
\begin{equation}\label{Lax pair1}
\psi_x=M\psi,
\end{equation}
\begin{equation}\label{Lax pair2}
\psi_t=(V_3\lambda^{3}+V_2\lambda^{2}+V_1\lambda+V_0)\psi=N\psi,
\end{equation}
with
\begin{equation*}
\psi=\left(
\begin{array}{c}
\phi_1\\\phi_2
\end{array}\right),
M=\left(
\begin{array}{cc}
-i\lambda & q\\
r & i\lambda
\end{array}\right),
V_3=\left(
\begin{array}{cc}
-4i & 0\\0 & 4i
\end{array}\right),
\end{equation*}

\begin{equation*}
V_2=\left(
\begin{array}{cc}
0 & 4q\\4\gamma & 0
\end{array}\right),
V_1=\left(
\begin{array}{cc}
-2iqr & 2iq_x\\-2ir_x & 2iqr
\end{array}\right),
\end{equation*}

\begin{equation*}
V_0=\left(
\begin{array}{cc}
-qr_x+q_xr & -q_{xx}+2q^{2}r\\-r_{xx}+2qr^{2} & qr_x-q_xr
\end{array}\right).
\end{equation*}
Here, $\lambda$ is an arbitrary complex spectral parameter or also called
eigenvalue, and $\psi$ is  the eigenfunction corresponding to
$\lambda$ of the complex mKdV equation. From the compatibility
condition $M_t-N_x+[M,N]=0$, one can easily obtain the coupled
equations (\ref{coupledmkdvq}) and (\ref{coupledmkdvr}).
Furthermore, set $T$ be a gauge transformation by
\begin{equation}\label{gauge}
\psi^{[1]}=T\psi, q\rightarrow q^{[1]}, r\rightarrow r^{[1]},
\end{equation}
and
\begin{equation}\label{x} \psi^{[1]}_{x}=M^{[1]}\psi^{[1]}, \
M^{[1]}=(T_x+TM)T^{-1},
\end{equation}
\begin{equation}\label{t}
\psi^{[1]}_{t}=N^{[1]}\psi^{[1]}, \ N^{[1]}=(T_t+TN)T^{-1}.
\end{equation}
Here, $M^{[1]}=M(\rightarrow q^{[1]},r\rightarrow r^{[1]}),
N^{[1]}=N(q \rightarrow q^{[1]},r\rightarrow r^{[1]})$. By
cross-differentiating (\ref{x}) and (\ref{t}), we obtain
\begin{equation}
M^{[1]}_{t}-N^{[1]}_{x}+[M^{[1]},N^{[1]}]=T(M_{t}-N_{x}+[M,
N])T^{-1}.
\end{equation}
This implies that, in order to prove that the mKdV equation is invariant
under the gauge transformation (\ref{gauge}), it is important to
look for determine the $T$ such that $M^{[1]}$, $N^{[1]}$ have the same
forms as $M$, $N$. Meanwhile, the seed solutions ($q$, $r$) in
spectral matrixes $M$, $N$ are mapped into the new  solutions
($q^{[1]}$, $r^{[1]}$) in terms of transformed spectral matrixes
$M^{[1]}$, $N^{[1]}$.

Recently, using the generalized Darboux transformation, $n$th-order
rogue wave solutions for the complex mKdV equation have been
proposed in e.g.\cite{Zhaqilao}. However, in our work, we shall
systematically analyze the evolution of the different patterns of
higher-order rogue waves by suitably choosing the parameters in the
rational solutions. In addition, it is worth to note that the
obtained results are in agreement with our recently published
developments about the method of generating higher-order rogue waves
\cite{liwuwanghe2013,hezhangwangpf2012}.
\\
\\
{\bf 2.1 One-fold Darboux Transformation}
\\
\\
From the knowledge of the known form of the DT for the nonlinear
Schr\"odinger equation \cite{GN,teacherli,Matveev,Gu1,Gu2,he2006},
we assume that a trial  Darboux matrix $T$ in eq. (\ref{gauge}) has
the following form
\begin{equation}
T=T(\lambda)=\left(
\begin{array}{cc}
a_1 & b_1\\c_1 & d_1
\end{array}\right)\lambda+\left(\begin{array}{cc}
a_0 & b_0\\c_0 & d_0
\end{array}\right),
\end{equation}
where $a_0$, $b_0$, $c_0$, $d_0$, $a_1$, $b_1$, $c_1$, $d_1$ are
functions of $x$ and $t$. From
\begin{equation}\label{Tx}
T_{x}+TM=M^{[1]}T,
\end{equation}
by comparing the coefficients of $\lambda^{j}$, $j=2,1,0$, it yields
\begin{eqnarray}\label{xpart}
&\mbox{\hspace{-1cm}}&\lambda^{2}: b_1=0,\
c_1=0,\nonumber\label{tlambda2}\\
&&\lambda^{1}:a_{1x}=0,\ -2ib_0+q_1d_1-qa_1=0,\nonumber\\
&&\qquad d_{1x}=0,\ -rd_1+r_1a_1+2ic_0=0,\nonumber\label{tlambda1}\\
&&\lambda^{0}:  q_1c_0-a_{0x}-rb_0=0,\ -b_{0x}+q_1d_0-qa_0=0,\nonumber\\
&&\qquad r_1a_0-c_{0x}-rd_0=0,\
r_1b_0-d_{0x}-qc_0=0.\label{tlambda0}
\end{eqnarray}
From the coefficients of $\lambda^{1}$, we conclude that $a_1$ and
$d_1$ are functions of $t$ only. Similarly, from
\begin{equation}\label{Tt}
T_{t}+TN=N^{[1]}T,
\end{equation}
and by comparing the coefficients of $\lambda^{j}$, $j=3,2,1,0$, we
obtain the following set of equations
\begin{eqnarray}\label{tpart}
&\mbox{\hspace{-1cm}}&\lambda^{3}: q_1d_1-qa_1-2ib_0=0,\
r_1a_1-rd_1+2ic_0=0,\nonumber\label{tlambda3}\\
&&\lambda^{2}: -q_1r_1a_1i+2q_1c_0+a_1qri-2rb_0=0,\
-a_1q_{x}i+2q_1d_0-2qa_0+q_{1x}d_1i=0,\nonumber\\
&&\qquad 2r_1a_0-r_{1x}a_1i-2rd_0+d_1r_xi=0,\
q_1r_1d_1i+2r_1b_0-2qc_0-d_1qri=0,\nonumber\label{tlambda2}\\
&&\lambda^{1}: -a_{1t}+r_1q_{1x}a_1+a_1qr_x-a_1rq_x+2iq_{1x}c+2ib_0r_x-2iq_1r_1a_0+2ia_0qr-q_1r_{1x}a_1=0,\nonumber\\
&&\qquad -2iq_1r_1b_0-2ib_0qr+a_1q_{xx}+2iq_{1x}d_{0}-2ia_0q_x-2a_1q^{2}r-q_{1xx}d_1+2q_1^{2}r_1d_1=0,\nonumber\\
&&\qquad -r_{1xx}a_1+2iq_1r_1c_0+2q_1r^{2}_1a_1+2iqrc_0-2ia_0r_{1x}+2id_0r_x-2d_1qr^{2}+d_1r_{xx}=0,\nonumber\\
&&\qquad -d_{1t}-2ic_0q_x+2iq_1r_1d_0-r_1q_{1x}d_1-2ir_{1x}b_0-d_1qr_x+d_1rq_x-2id_0qr+q_1r_{1x}d_1=0,\nonumber\label{tlambda1}\\
&&\lambda^{0}: -q_{1xx}c_0+b_0r_{xx}+2q^{2}_1r_1c_0+a_0qr_x-2b_0qr^{2}+r_1q_{1x}a_0-q_1r_{1x}a_0-a_0rq_x-a_{0t}=0,\nonumber\\
&&\qquad a_0q_{xx}-b_0qr_x-b_0q_1r_{1x}-2a_0q^{2}r+r_1q_{1x}b_0+2q^{2}_1r_1d_0+b_0rq_x-q_{1xx}d_0-b_{0t}=0,\nonumber\\
&&\qquad-r_{1xx}a_0+d_0r_{xx}-r_1q_{1x}c_0+2q_1r^{2}_1a_0+q_1r_{1x}c_0-c_0rq_x+c_0qr_x-2d_0qr^2-c_{0t}=0,\nonumber\\
&&\qquad-r_1q_{1x}d_0+q_1r_{1x}d_0-r_{1xx}b_0+2q_1r^{2}_1b_0+c_0q_{xx}+d_0rq_x-d_0qr_x-2c_0q^{2}r-d_{0t}=0.\label{tlambda0}
\end{eqnarray}
By making use of eq. (\ref{xpart}) and eq. (\ref{tpart}), one may obtain
$a_{1x}=0, \ d_{1x}=0,\ a_{1t}=0,\ d_{1t}=0$, which implies that $a_1$ and $d_1$ are
two constants.

 In order to obtain the non-trivial solutions of the complex mKdV equation,
we provide the Darboux transformation under the condition $a_1=1,\ d_1=1$.
Without loss of generality, and based on eqs. (\ref{xpart}) and (\ref{tpart}),
we observe that the Darboux matrix T admits the following form
 \begin{equation}
T=T(\lambda)=\left(
\begin{array}{cc}
1 & 0\\0 & 1
\end{array}\right)\lambda+\left(\begin{array}{cc}
a_0 & b_0\\c_0 & d_0
\end{array}\right).
\end{equation}
Here, $a_0$, $b_0$, $c_0$, $d_0$ are functions of $x$ and $t$, which could
be expressed by two eigenfunctions corresponding to  $\lambda_1$
and $\lambda_2$. To begin with, we introduce $2n$ eigenfunctions
$\psi_j$ and $2n$ associated distinct eigenvalues $\lambda_j$ as
follows
 \begin{equation}
\psi_j=\left(
\begin{array}{cc}
\phi_{j1}\\ \phi_{j2}
\end{array}\right),\ j=1,2,......2n,\  \phi_{j1}=\phi_{1}(x,t,\lambda_j),
\  \phi_{j2}=\phi_{2}(x,t,\lambda_j).
\end{equation}
Note $\phi_1(x,t,\lambda)$ and $\phi_2(x,t,\lambda)$ are two
components of eigenfunction $\psi$ associated with $\lambda$ in
eqs. (\ref{Lax pair1}) and (\ref{Lax pair2}).  Here,  it is worthwhile
to note that since the eigenfunction
\begin{equation*} \psi_j=\left(
\begin{array}{cc}
\phi_{j1}\\ \phi_{j2}
\end{array}\right)
\end{equation*}
is the solution of the eigenvalue equations (\ref{Lax pair1}) and
(\ref{Lax pair2}) corresponding to $\lambda_j$, and the eigenfunction
\begin{equation*}
\psi'_j=\left(
\begin{array}{cc}
-\phi^{*}_{j2}\\ \phi^{*}_{j1}
\end{array}\right)
\end{equation*}
is also the solution of eqs. (\ref{Lax pair1}) and (\ref{Lax
pair2}) corresponding to $\lambda^{*}_{j}$, where $*$ denotes the
complex conjugate.

We assume from now on that even number
eigenfunctions and eigenvalues are given by odd ones as the following
rule ($j=1,2,\dots,n$):
\begin{eqnarray}\label{eigenfunctionchoice}
\lambda_{2j}=\lambda^*_{2j-1},
\phi_{2j,1}=-\phi^*_{2j-1,2}(\lambda_{2j-1}),
\phi_{2j,2}=\phi^*_{2j-1,1}(\lambda_{2j-1}).
\end{eqnarray}
For convenience and simplicity of our mathematical manipulations, we
propose the following theorems:

\noindent {\bf Theorem 1.} {\sl The elements of a one-fold Darboux
matrix are presented with the eigenfunction $\psi_1$ corresponding
to the eigenvalue $\lambda_1$ as follows
\begin{eqnarray}
a_{0}=-\frac{1}{\Delta_2}\left|\begin{array}{cc}
 \lambda_1\phi_{11}&\phi_{12} \\
\lambda_2\phi_{21}&\phi_{22}
 \end{array}\right|,\ \
b_{0}=\frac{1}{\Delta_2}\left|\begin{array}{cc}
 \lambda_1\phi_{11}&\phi_{11} \\
\lambda_2\phi_{21}&\phi_{21}
 \end{array}\right|,\qquad \qquad \qquad \nonumber\\
c_{0}=\frac{1}{\Delta_2}\left|\begin{array}{cc}
 \phi_{12}&\lambda_1\phi_{12} \\
\phi_{22}&\lambda_2\phi_{22}
 \end{array}\right|,\ \
d_{0}=-\frac{1}{\Delta_2}\left|\begin{array}{cc}
 \phi_{11}&\lambda_1\phi_{12} \\
\phi_{21}&\lambda_2\phi_{22}
 \end{array}\right|,\qquad \qquad \qquad
 \label{DT1a0b0c0d0}\\
\Leftrightarrow T_1(\lambda;\lambda_1)=\left(
\begin{array}{cc}
\lambda-\frac{1}{\Delta_2}\left|\begin{array}{cc}
 \lambda_1\phi_{11}&\phi_{12} \\
\lambda_2\phi_{21}&\phi_{22}
 \end{array}\right|&
\frac{1}{\Delta_2}\left|\begin{array}{cc}
 \lambda_1\phi_{11}&\phi_{11} \\
\lambda_2\phi_{21}&\phi_{21}
 \end{array}\right|\\
\frac{1}{\Delta_2}\left|\begin{array}{cc}
 \phi_{12}&\lambda_1\phi_{12} \\
\phi_{22}&\lambda_2\phi_{22}
 \end{array}\right| &\lambda-\frac{1}{\Delta_2}\left|\begin{array}{cc}
 \phi_{11}&\lambda_1\phi_{12} \\
\phi_{21}&\lambda_2\phi_{22}
 \end{array}\right|  \\
\end{array}  \right),\label{DT1matrix}
\end{eqnarray}
with $\Delta_2=\left|\begin{array}{cc}
 \phi_{11}&\phi_{12} \\
\phi_{21}&\phi_{22}
 \end{array}\right|$,  and
 then the new solutions $q^{[1]}$ and $r^{[1]}$ are given
by
\begin{eqnarray}\label{t1q}
q^{[1]}=q+2i\frac{1}{\Delta_2}\left|\begin{array}{cc}
 \lambda_1\phi_{11}&\phi_{11} \\
\lambda_2\phi_{21}&\phi_{21}
 \end{array}\right|,\ \
r^{[1]}=r-2i\frac{1}{\Delta_2}\left|\begin{array}{cc}
 \phi_{12}&\lambda_1\phi_{12} \\
\phi_{22}&\lambda_2\phi_{22}
 \end{array}\right|,
\end{eqnarray}
and the new eigenfunction $\psi_j^{[1]}$ corresponding to
$\lambda_j$ is
\begin{equation}
\psi^{[1]}_j=T_1(\lambda;\lambda_1)|_{\lambda=\lambda_j}\psi_j.
\end{equation}\vspace{-0.5cm}
}\\
\noindent{\bf Proof.} Note that $b_{1}=c_{1}=0$, ${a_{1}}_{x}=0$ and
${d_{1}}_{x}=0$ is derived from the functional form of $x$, then
${a_{1}}_{t}=0$ and ${d_{1}}_{t}=0$ is derived from the functional
form of $t$. So, $a_{1}$ and $d_{1} $ are arbitrary constants, and
hence, we let $a_{1}=d_{1}=1$ for simplicity for later calculations.
By transformation defined by eq. (\ref{xpart}) and eq. (\ref{tpart}), new
solutions are given by
 \begin{eqnarray} \label{TT1}
q_{1}=q+2ib_{0},r_{1}=r-2ic_{0}.
\end{eqnarray}
By making use of the general property of the DT, i.e.,
 $T_{1}(\lambda;\lambda_{j})|_{\lambda=\lambda_1}\psi_{j}=0, j=1,2$, after some manipulations,  eq. (\ref{DT1a0b0c0d0}) is obtained.
Next, substituting $(a_0,b_0,c_0,d_0)$ given in
eq. (\ref{DT1a0b0c0d0}) into eq. (\ref{TT1}), the new solutions are
given as in eq. (\ref{t1q}). Furthermore, by using the explicit
matrix representation eq. (\ref{DT1matrix}) of $T_1$, then
$\psi^{[1]}_j(j\geq 3)$ is given by
$\psi^{[1]}_j=T_1(\lambda;\lambda_1)|_{\lambda=\lambda_j} \psi_j$. $\square$\\
It is trivial to confirm $q^{[1]}=-(r^{[1]})^*$ by using the special
choice on $\psi_2$ and $\lambda_2$ in
eq. (\ref{eigenfunctionchoice}). This means $q^{[1]}$ generates a new
solution of the complex mKdV from a seed solution $q$. Note that
$\psi^{[1]}_j=0$ for $j=1,2$.
\\
\\
{\bf  2.2 n-fold Darboux transformation}
\\
\\
By $n$-times iteration of the one-fold DT $T_1$, we obtain $n$-fold DT
$T_n$ of the complex mKdV equation with the special choice on
$\lambda_{2j}$ and $\psi_{2j}$ in eq. (\ref{eigenfunctionchoice}). To
save space, we omit the tedious calculation of  $T_n$ and its
determinant representation. Under the above conditions, the reduction
condition $q^{[n]}=-(r^{[n]})^*$ is preserved by $T_n$, so we just
give $q^{[n]}$ in the following theorem.
\\
\\
\noindent {\bf Theorem 2.} {\sl  Under the choice of
eq. (\ref{eigenfunctionchoice}), the n-fold DT $T_n$ generates a new
solution of the complex mKdV equation from a seed solution $q$ as
\begin{eqnarray}\label{nT}
q^{[n]}=q-2i\frac{N_{2n}}{D_{2n}},
\end{eqnarray}
where
\begin{equation}\label{numer}
N_{2n}=\begin{vmatrix}
\phi_{11}&\phi_{12}&\lambda_1\phi_{11}&\lambda_{1}\phi_{12}&\ldots&\lambda_{1}^{n-1}\phi_{11}&\lambda_{1}^{n}\phi_{11}\\
\phi_{21}&\phi_{22}&\lambda_2\phi_{21}&\lambda_{2}\phi_{22}&\ldots&\lambda_{2}^{n-1}\phi_{21}&\lambda_{2}^{n}\phi_{21}\\
\phi_{31}&\phi_{32}&\lambda_3\phi_{31}&\lambda_{3}\phi_{32}&\ldots&\lambda_{3}^{n-1}\phi_{31}&\lambda_{3}^{n}\phi_{31}\\
\phi_{41}&\phi_{42}&\lambda_4\phi_{41}&\lambda_{4}\phi_{42}&\ldots&\lambda_{4}^{n-1}\phi_{41}&\lambda_{4}^{n}\phi_{41}\\
\vdots&\vdots&\vdots&\vdots&\vdots&\vdots&\vdots\\
\phi_{2n1}&\phi_{2n2}&\lambda_{2n}\phi_{2n1}&\lambda_{2n}\phi_{2n2}&\ldots&\lambda_{2n}^{n-1}\phi_{2n1}&\lambda_{2n}^{n}\phi_{2n1}\nonumber\\
\end{vmatrix},
\end{equation}
\begin{equation}\label{denom}
D_{2n}=\begin{vmatrix}
\phi_{11}&\phi_{12}&\lambda_1\phi_{11}&\lambda_{1}\phi_{12}&\ldots&\lambda_{1}^{n-1}\phi_{11}&\lambda_{1}^{n-1}\phi_{12}\\
\phi_{21}&\phi_{22}&\lambda_2\phi_{21}&\lambda_{2}\phi_{22}&\ldots&\lambda_{2}^{n-1}\phi_{21}&\lambda_{2}^{n-1}\phi_{22}\\
\phi_{31}&\phi_{32}&\lambda_3\phi_{31}&\lambda_{3}\phi_{32}&\ldots&\lambda_{3}^{n-1}\phi_{31}&\lambda_{3}^{n-1}\phi_{32}\\
\phi_{41}&\phi_{42}&\lambda_4\phi_{41}&\lambda_{4}\phi_{42}&\ldots&\lambda_{4}^{n-1}\phi_{41}&\lambda_{4}^{n-1}\phi_{42}\\
\vdots&\vdots&\vdots&\vdots&\vdots&\vdots&\vdots\\
\phi_{2n1}&\phi_{2n2}&\lambda_{2n}\phi_{2n1}&\lambda_{2n}\phi_{2n2}&\ldots&\lambda_{2n}^{n-1}\phi_{2n1}&\lambda_{2n}^{n-1}\phi_{2n2}\nonumber\\
\end{vmatrix}.
\end{equation}}

 By making use of Theorem 2 with a suitable seed solution,
we can generate the multi-solitons, multi-breathers, and multi-rogue
waves of the complex mKdV equation.  As the multi-soliton and
multi-breather solutions are well-known and completely explored for
the complex mKdV equation, next, we shall concentrate mainly on the
systematic construction of the higher-order rogue waves from the double
degeneration \cite{hezhangwangpf2012} of the DT. Though the
construction of higher-order rogue wave solutions is quite
cumbersome, one can still validate the correctness of these solutions
with the help of modern computer tools such as a simple symbolic
calculation or equivalent, and also by a direct numerical computation.

\section{Higher-order rogue waves}
In this section, starting with a non-zero seed $q=ce^{i\rho},\
\rho=ax+bt,\ b=a^3-6ac^2$, $a,b,c\in \mathbb{R}$, we shall present
higher-order rogue waves of the complex mKdV equation. If $a=0$, $q=c$ a constant, which is just a seed solution to generate
soliton. So,  in this paper, we choose $a \neq 0$.    By using
the principle of superposition of the linear differential
equations, then, the new eigenfunctions corresponding to $\lambda_j$
can be provided by
\begin{equation}\label{neweigenfunction}
\psi_j=\left(\begin{array}{c}
d_1 ce^{i[(\frac{1}{2}a+c_1)x+(\frac{1}{2}b+2c_1c_2)t]} + d_2 i(\frac{1}{2}a+\lambda_j+c_1)e^{i[(-\frac{1}{2}a+c_1)x+(-\frac{1}{2}b+2c_1c_2)t]}\\
d_1
i(\frac{1}{2}a+\lambda_j+c_1)e^{-i[(-\frac{1}{2}a+c_1)x+(-\frac{1}{2}b+2c_1c_2)t]}+
d_2 ce^{-i[(\frac{1}{2}a+c_1)x+(\frac{1}{2}b+2c_1c_2)t]}
\end{array}\right)
\end{equation}
with
\begin{equation}
d_1=e^{ic_1(s_0+s_1\varepsilon+s_2\varepsilon^{2}+\ldots+s_{n-1}\varepsilon^{n-1})},\
d_2=e^{-ic_1(s_0+s_1\varepsilon+s_2\varepsilon^{2}+\ldots+s_{n-1}\varepsilon^{n-1})},
\nonumber
\end{equation}
\begin{equation}
c_1=\frac{1}{2}\sqrt{a^2+4c^2+4\lambda_ja+4\lambda_j^2},\ \
c_2=2\lambda^{2}_j-c^2+\frac{1}{2}a^2-\lambda_j a.
\end{equation}
Here, $s_i\in \mathcal{C}(i=0,1,2,\cdots, n-1), a,b,c \in \mathbb{R}$
are the arbitrary constants, $\varepsilon$ is an infinitesimal
parameter.

We are now in a position to consider the double degeneration of
$q^{[n]}$ to obtain higher-order rogue wave as in our earlier
investigations \cite{hezhangwangpf2012}. It is trivial to check that
$\psi_j(\lambda_0)=0$ in eq. (\ref{neweigenfunction}), which means that
these eigenfunctions are degenerate at $\lambda_0=ic-\frac{a}{2}$.
Setting $\lambda_{2j-1}\rightarrow \lambda_0$ and substituting
$\psi_{2j-1}(j=1,2,\cdots,n)$ defined by eq. (\ref{neweigenfunction})
back into eq. (\ref{nT}), the double degeneration, i.e. eigenvalue
and eigenfunction degeneration, occurs in $q^{[n]}$. Next,
$q^{[n]}$ now becomes an indeterminate form $\frac{0}{0}$. Set
$\lambda_{2j-1}=\lambda_0+\epsilon$ and set $\psi_{2j-1}$ be given
by eq. (\ref{neweigenfunction}), we obtain $n$-th order rogue wave
solutions by higher-order Taylor expansion of $q^{[n]}$ with respect
to $\epsilon$.\\

\noindent {\bf Theorem 3.} {\sl An n-fold degenerate DT with a given
eigenvalue $\lambda_0$ is realized in the degenerate limit
$\lambda_j \rightarrow \lambda_0$ of $T_n$. This degenerate n-fold
DT yields a new solution $q^{[n]}$ of the mKdV equation starting
with the seed solution $q$,where
\begin{eqnarray}
q^{[n]}(x,t;\lambda_0)=q-2i\frac{N'_{2n}}{D'_{2n}},
\end{eqnarray}
with
\begin{eqnarray*}
D'_{2n}=(\frac{\partial^{n_{i}}}{\partial\varepsilon^{n_{i}}}
\mid_{\varepsilon=0}(D_{2n})_{ij}(\lambda_0+\epsilon))_{2n\times2n},\\
N'_{2n}=(\frac{\partial^{n_{i}}}{\partial\varepsilon^{n_{i}}}
\mid_{\varepsilon=0}(N_{2n})_{ij}(\lambda_0+\epsilon))_{2n\times2n}.
\end{eqnarray*}
Here, $n_{i}=[\frac{i+1}{2}]$, $[i]$ denotes the floor function of $i$.}

 In the following, to avoid the tedious mathematical steps we encountered,
we only present the expressions of the 1st,  2nd and 3rd  order rogue waves
by using Theorem 3.  In each case, the solution $q^{[n]}$
describes the envelope of the rogue wave, and its square modulus
contains information such as e.g. wave evolution above water
surface, or the intensity of few-cycle optical wave, etc.

Firstly,  we set $n=1$, $D_2$ and $N_2$ take the form of $2\times 2$ determinants. By using the 1st-order Taylor expansion with respect to $\epsilon$ in terms of elements of $D_2$ and $N_2$ through $\lambda_1=\lambda_0+\epsilon$, we determined $N'_2$ and $D'_2$
by equating the coefficient of $\sqrt{\epsilon}$, and then obtained the explicit expression
for the 1st-order rogue wave as
\begin{equation}
q^{[1]}=-ce^{ia(x+t(a^2-6c^2))}\frac{A+48ic^2ta-3}{A+1},
\end{equation}
with $A=24ta^2c^2x+24ta^2c^2s_0+36t^2a^4c^2
-48c^4tx-48c^4ts_0+8c^2xs_0 +144c^6t^2+4c^2x^2+4c^2s_0^2.$\\
Its evolution is presented in Figure 1 (left) with the condition
$d_1=d_2=1$ and the Taylor expansion at
$\lambda_0=ic-\frac{a}{2}+\varepsilon$, in order to compare this with
higher-order rogue waves.  It is trivial to find that $|q^{[1]}|^2=c^2$ when
$x \rightarrow \infty$  and $t \rightarrow \infty$. This means that the asymptotic plane of $|q^{[1]}|^2$ has the
height $c^2$.  Particularly, let $a=0$ and $s_0=0$, $|q^{[1]}|^2$ is a soliton propagating along a line $x=6c^2t$ with a non-vanishing
boundary. Set $c=-1,a=\sqrt{6},s_0=0$ and $t\rightarrow t/2$, then $q^{[1]}$ gives $u[2]$ of
ref.\cite{Zhaqilao}.

When $n=2$, we construct the 2nd-order rogue waves under the
assumption $d_1=e^{ic_1(s_0+s_1\varepsilon)},\
d_2=e^{-ic_1(s_0+s_1\varepsilon)},s_0=0$ from Theorem 3.  An
explicit form of $q^{[2]}$ is constructed as
\begin{equation}\label{2ndrw}
q^{[2]}=ce^{ia(x+t(a^2-6c^2))}\frac{B}{C}.
\end{equation}
Here, $B$ and $C$ are  two degree 6 polynomials in $x$ and $t$,
which are given in appendix A. From Figure 1 (right), one finds that under the assumption $d_1=1,d_2=1$,
or equivalently $s_0=s_1=0$,  the second-order rational solution
admits a single high maximum at the origin. By suitably adjusting
the parameter $a$ one could control the decaying rate
of the profile in the $(x,t)$-plane. This is a fundamental
pattern. Furthermore, as is shown in Figure 6, when taking
$d_1\neq1$ and $d_2\neq1$, the large peak of the 2nd rogue wave is
completely separated and forms a set of three first-order
rational solution for sufficiently large $s_1$ meanwhile $s_0=0$, and
actually forms an equilateral triangle.

When $n=3$, and set
$d_1=e^{ic_1(s_0+s_1\varepsilon+s_2\varepsilon^2)}$ and $\
d_2=e^{-ic_1(s_0+s_1\varepsilon+s_2\varepsilon^2)}$, then Theorem 3
yields an explicit formula of the 3rd rogue wave with parameters
$a,c, s_0,s_1,s_2$. Set $a=1.5, c=1, s_0=s_1=s_2=0$, we have
\begin{equation}\label{3rd}
q^{[3]}= \dfrac{L_1}{L_2}e^{i(\frac{3}{2}x-\frac{45}{8}t)}.
\end{equation}
Here, $L_1$ and $L_2$ are  two degree 12 polynomials in $x$ and $t$,
which are given in appendix B. This is the fundamental pattern of the
3rd-order rogue wave, which is plotted in Figure 2(left) with a different
value of $a$.

In general, Theorem 3 provides an efficient tool to produce
analytical forms of higher-order rogue waves of the complex mKdV
equation. Actually, we have also constructed the analytical formulas
for 4th, 5th and 6th -order rogue waves. However, because of their
long expressions  describing these solutions, we do not present
them here but would provide upon request. The validity of
all these higher-order rogue waves has been verified by
symbolic computation. According to the explicit formulae of the $n$th-order rogue waves
under fundamental patterns, we find that their maximum amplitude is $(2n+1)^2c^2(n=1,2,3,4,5,6)$ by setting
$x=0$ and $t=0$ in $|q^{[n]}|^2$, and the height of the asymptotic plane is $c^2$,
 which is the same as that of the  rogue wave of the NLS equation.  This fact can be easily verified through Figs.(1-3).
All figures in this paper are plotted based on
these explicit analytical formulas of the solutions. Once the
explicit analytical higher-order rogue waves are known, our next aim
is to generate and understand underlying the dynamics of the obtained
different patterns by suitably selecting the value of $s_i$ .

\section{Results and discussion}

The above discussion is a clear manifestation of the evolution of the
higher-order rogue waves from the Taylor expansion of the degenerate
 breather solutions. A brief discussion about the generating mechanism
 of higher-order rogue waves from the nonlinear evolution equation has
 already been reported by \cite{hezhangwangpf2012}. For our
purpose now, we customize our discussion only up to 6th-order rogue
waves, since higher-order rogue waves are difficult to construct owing
to the extreme complexity and tedious mathematical calculations.
It is quite obvious from our numerical analysis that the choice of
parameters such as $d_1$ and $d_2$ actually do generate three
different basic patterns of rogue wave solutions.
Let us discuss these patterns now.\\

Fundamental patterns:  When, e.g. $d_1 = d_2 = 1$, or
equivalently $s_i=0\ (i=0,1,2,\cdots,n-1)$ in $q^{[n]}$, the rational
solutions of any order $n$ have a similar structure. In addition,
there are $\frac{n(n+1)}{2} - 1$ local maxima on each side of the
line at $t = 0$. Starting from $\infty$, before the central optimum
high amplitude, there is a sequence of peaks with gradual increase
in height. Here, one can observe that the number of first peaks is $n$,
then there is a row of $n - 1$ symmetric peaks with respect to time
$t$ as shown in Figs. (1-3) for 6 rogue waves.

There are only two parameters $a$ and $c$ in the explicit forms of the rogue waves under
fundamental patterns. It is a challenge problem to illustrate analytically the role of $a$ and $c$ in the
control of the profile for the higher-order RWs due to the extreme complexity of the explicit forms of
the nth-order RWs ($n\geq 2$). So, we only study this problem for the first-order RW\ $|q^{[1]}|^2$.
To this end, we introduce a method, i.e., the contour line method, to analyze  the
contour profile of the red bright spots in the density plot of Fig. 4, which intuitively shows the localization
characters such as length and width of the RW. On the background plane with  height $c^2$,  a contour line of
$|q^{[1]}|^2$ with $c=1$ is a hyperbola
\begin{equation}\label{contourlineonasym}
x^2-12tx+6ta^2x+36t^2-72t^2a^2+9t^2a^4-\frac{1}{4}=0,
\end{equation}
which has  two asymptotes
\begin{eqnarray}\label{asymtptoticlines}
& l_1: x=(6-3a^2-6a)t,\ \ \  l_2: x=(6-3a^2+6a)t,
\end{eqnarray}
and two non-orthogonal axes:
\begin{eqnarray}\label{twoaxes}
&{\rm major\ axis}: t= 0,\ \ \ {\rm imaginary\ axis} (l_3): x=(6-3a^2)t.
\end{eqnarray}
There are two fixed vertices: $P_1=(0,0.50), P_2=(0,-0.5)$ on ($t,x$) plane of all value of $a$.
Here, $l_3$ is also a median of one triangle composed of above two asymptotes and a parallel line
 of $x$-axis except $t=0$. We combine the density plots and the above three lines in Fig.4 with different values of $a$.
 At height $c^2+1$, a contour line of
$|q^{[1]|^2}$ with $c=1$ is given by a quartic polynomial
\begin{eqnarray}\label{contourlinehigherasym}
&x^4+(12a^2-24)tx^3+(\frac{5}{2}+(216-144a^2+54a^4)t^2)x^2+((-864+432a^2-216a^4+108a^6)t^3 \nonumber \\
&+(15a^2-30)t)x+(1296+648a^4+81a^8)t^4+(90-144a^2+( \frac{45}{2})a^4)t^2- \frac{7}{16}=0,
\end{eqnarray}
which has two end points $P_3=(-\frac{\sqrt{7}}{12a}, \frac{(-2+a^2)\sqrt{7}}{4a})$
 and  $P_4=(\frac{\sqrt{7}}{12a},  -\frac{(-2+a^2)\sqrt{7}}{4a})$ along $t$-direction.
Moreover, there are  two fixed points expressed by $P_5=(0, \ \ \frac{\sqrt{-1-4c^2+4\sqrt{c^2(c^2+1)}}}{2c}|_{c=1})=(0,0.41)$, \\ $P_6=(0, -\frac{\sqrt{-1-4c^2+4\sqrt{c^2(c^2+1)}}}{2c}|_{c=1})=(0,-0.41)$ on ($t,x$) plane of all value of $a$. At height $\frac{c^2}{2}$,
a contour line of $|q^{[1]}|^2$ with $c=1$ is also given by a quartic polynomial
\begin{eqnarray}\label{contourlinelowerasym}
&x^4+(12a^2-24)tx^3+((216-144a^2+54a^4)t^2-\frac{7}{2})x^2+((-864+432a^2-216a^4+108a^6)t^3 \nonumber \\
&+(-21a^2+42)t)x+(1296+648a^4+81a^8)t^4+(-126+288a^2-\frac{63}{2}a^4)t^2+ \frac{17}{16}=0,
\end{eqnarray}
which  is defined on interval [-$ \frac{1}{12a},\frac{1}{12a} $] of $t$.
For this contour line, there are  four fixed points: $(0,1.78), (0,0.58), (0,-0.58), (0,-1.78)$ on
($t,x$) plane of all value of $a$. Two centers of valleys of $|q^{[1]}|^2$ given by $P_7=(0,\frac{\sqrt{3}}{2c}), P_8=(0,-\frac{\sqrt{3}}{2c})$,
which are  independent with the value of $a$. Fig.5 are plotted for above contour lines with different values of $a=1.5,2,2.5$.

Based on the above analytical results,  we could define the length and width of the rogue wave, which are two crucial characters of
a doubly localized wave-RW. Because the contour line of RW on the  background plane is not a closed curve,
so we can not define a length for RW on this plane. However, set $d$ be a positive constant, the contour line at height $c^2+d$
is closed. Here $c^2+d<9c^2$, or equivalently $d<\sqrt{8c^2}$ because the max amplitude of the first-order RW ($|q^{[1]}|^2$) is $9c^2$.
Without loss of generality, and considering a recognizable height from the asymptotic plane, we set $d=1$ as before. We can use the length of
the area surrounded by the contour line at height $c^2+1$ as the length of the first-order RW. The length-direction is defined by ${\sl l_3}$, the width-direction is orthogonal to it.  The reasons for this choice are :
1) $l_3$ passes through $P_3$ and $P_4$; 2) $l_3$ is parallel to the
tangent line of  hyperbola at two vertices; 3) $l_3$ is parallel to the
tangent line  of the contour line at $P_5$ and $P_6$.  Let $k_3$ be the slope of $l_3$.  So the length of the first-order RW is
the distance of $P_3$ and $P_4$, i.e.,
\begin{equation}
d_L=\frac{\sqrt{7}}{6a}\sqrt{1+(k_3)^2}=\frac{\sqrt{7}}{6a}\sqrt{1+(6-3a^2)^2}.
\end{equation}
The width is defined as the projection of line segment $P_7P_8$ at width-direction, which is expressed by
\begin{equation}
d_W=\frac{\sqrt{3}}{\sqrt{1+(k_3)^2}}=\frac{\sqrt{3}}{\sqrt{1+(6-3a^2)^2}}.
\end{equation}
$d_L$ and $d_W$ are plotted in figure 6 with fixed $c=1$, which shows that the length is decreased with  $a$  when $a \in (0, \sqrt{2})$
and is increased with $a$ when $a> \sqrt{2}$. However, the width has an opposite increasing or decreasing trend with respect to $a$.
When $a=\sqrt{2}$, the profile of first-order RW is parallel to the t-axis, then the length reaches to its minimum,
and the width reaches to its maximum.  This is the first role of $a$ in the control of the
RW. Moreover, we know from $k_3$ that the increase of $a$ results in the rotation of RW in the clockwise direction.
This is the second role of $a$.

In above discussion for the role of $a$, we have set $c=1$. If $c\not= 1$, it is a more interesting and complicated case, which can be studied
as above by using contour line method. To save the space, we shall provide corresponding results without explanation, which can be done by
a similar way as above.  Under this case,
two asymptotes of the contour line of  first-order RW $|q^{[1]}|^2$ at height $c^2$
\begin{eqnarray}\label{twoaxesc}
&{\rm major\ axis}: t= 0,\ \ \ {\rm imaginary\ axis (cl_3)}: x=(6c^2-3a^2)t.
\end{eqnarray}
In other word, slope is $k_{3c}=6c^2-3a^2$.   Two vertices of the hyperbola are  $P_1=(0,\frac{1}{2c})$ and $P_2=(0,-\frac{1}{2c})$ on ($t,x$)
 plane. For contour line at height $c^2+1$, two end points are
 $P_3=(-\frac{\sqrt{8c^2-1}}{12ac^2}, \frac{(-2c^2+a^2)\sqrt{8c^2-1}}{4ac^2})$
 and  $P_4=(\frac{\sqrt{8c^2-1}}{12ac^2}, -\frac{(-2c^2+a^2)\sqrt{8c^2-1}}{4ac^2})$ along $t$ direction.
 So the length of the first-order  RW is
 \begin{equation}
 d_{cL}=\frac{\sqrt{8c^2-1}}{6c^2a}\sqrt{1+9(-2c^2+a^2)^2},
 \end{equation}
 and the width of the first-order RW is
 \begin{equation}
 d_{cW}=\frac{\sqrt{3}}{c}\frac{1}{\sqrt{1+9(-2c^2+a^2)^2}},
 \end{equation}
 which are plotted in Figure 7. These pictured show visually the role of $a$ and $c$ in the  control of the first-order RW.
 For a given value of $a$, $d_{cL}$ has two extreme points with respect to $c$. However, for a given value of $c$,  $d_{cL}$ has one extreme point
 with respect to $a$.  The slope  $k_{3c}$ shows that the increasing of $a$ and $c$  results in the rotation of the first-order RW with different direction. Note that $a=\sqrt{2}c$ is a line of points for extreme value. Under this condition, the profile of first-order RW is parallel to
 the t-axis,  and the minimum of the length is $\frac{\sqrt{4a^2-1}}{3a^3}$, the maximum of the width is $\frac{\sqrt{3}}{c}.$

Triangular patterns: The triangular structure can be obtained by
choosing the first non-trivial coefficient $s_1 \gg 1$, while the
rest of the values are assumed to be zero. It can be seen from
Figs. (8-10) that the $n$-th order rational solutions have
$\frac{n(n+1)}{2}$  peaks of equal height with a structure of
equilateral triangular type having $n$ peaks at each edge.

Ring patterns: One can observe the ring structure/pattern when $n
\geq 3$ and the principle coefficient for $n$-th order rational
solution when $s_{n-1}\gg 1$, while the remaining coefficients $s_i$
are all zero. The rational solutions consist of the outer circular
shell of $2n-1$ first-order rational solutions, while the center
is an order $(n-2)$ rational solution of the fundamental patterns as
portrayed in Figs. (11-12).  The center order-($n-2$) rogue wave can be
decomposed further into different lower-order patterns according to
the ($n-2$)-reduction rule of order by setting one of
$s_i\ (i=0,1,2,\dots,n-3)$ to non-zero, which are plotted in
Figs. (13-15). For example, the center order-4 rogue wave of the $6$-th
rogue wave has a fundamental pattern (Figure 12 (right)), a ring plus a
fundamental pattern (Figure 14 (right)) or triangular pattern ((Figure 15 (right))) of
2nd-order rogue wave,  a triangular pattern (Figure 15 (left)).   We call these forms as  standard decomposition of
 the rogue wave. This structure is similar to the so-called ``wave clusters'' as reported
in\cite{akhmedievhigherorderrw2011}.

Another interesting fact worth mentioning here is that the profiles
of higher-order rogue wave are actually a complicated
combination of above three basic patterns: ``fundamental" pattern,
``ring'' pattern and ``triangular'' pattern, which can provide further
interesting patterns of the rogue waves. This can be achieved  by
suitably selecting the different values of $s_i$. In particular, one
can generate multi-ring structures but these rings do not possess
2n-1 peaks and also do not satisfy the rule of (n-2)-reduction of
order as mentioned earlier in the case of ring pattern formation.
Thus,  we call these formations as non-standard decomposition of the rogue waves.
Figures (16-20) represent a few examples of this kind of special ring
structures. One common feature, which we observed from these examples,
is the appearance of at least two ring patterns with the same number of
peaks. It should be noted that the center-most profile of Figure 18 (left)
is a fundamental pattern of a 2nd-order rogue wave, which clearly
shows that Figure 18 (left) is not a complete decomposition of the $6$-th
order rogue wave. On the other hand, Figure 18 (right) presents the
complete decomposition. To arrive to a  better understanding of the non-standard
decomposition, we provide the distribution of peaks in Table 1. From
Figure 19 (left), it can also be observed the occurrence of two ring
patterns with a single inner peak, however, the inner ring consists of $5$
triangular patterns. So, the distribution of peaks is $5+3\times
5+1$.

In spite of having different structures, all types of rogue wave
solutions possess certain commonality as follows: The total number of
peaks admitted by $n$-th order solutions is $\frac{n(n+1)}{2}$ in terms of
a complete decomposition pattern. These structures actually depend on
the choice of the free parameters. Among all parameters, the principle
coefficient $s_{n-1}$ is accountable for the formation of a ring
structure. The first non-trivial coefficient $s_1$ is responsible for
the evolution of a triangular structure.  Furthermore, to see difference between the RWs of the complex mKdV and the NLS clearly,
we use the first-order RW (\cite{hezhangwangpf2012}) of the NLS, i.e.,
\begin{equation}
q^{[1]}_{NLS}=c^2\frac{\tilde{A}-32c^2((-x+2at)^2-4c^2t^2)+8}{\tilde{A}},\ \ \tilde{A}=(4c^2x^2-16c^2xta+16t^2(c^4+c^2a^2)+1)^2.
\end{equation}
to calculate the contour lines at heights $c^2$ and $c^2+1$, and to calculate the length, width by the same procedure we have used
in complex mKdV.  Here the NLS equations is in the form of
\begin{equation}
iq_t+q_{xx}+2|q^2|q=0.
\end{equation}
Similar to the contour line method of the complex mKdV, we get the slope of the imaginary axis of the hyperbola formed by a contour line of the  $|q^{[1]}_{NLS}|^2$ on background plane with a height $c^2$:$k_{3cNLS}=2a$, and
the length of the RW
\begin{equation}
d_{cLNLS}=\frac{1}{2c^2}\sqrt{(-1+8c^2)(1+4a^2)},
\end{equation}
and the width of the RW
\begin{equation}
d_{cWNLS} =\frac{\sqrt{3}}{c\ \sqrt{1+4a^2}}.
\end{equation}
The dynamical evolution of the first-order RW $|q_{NLS}^{[1]}|^2$ of the NLS are plotted in Fig. 21, contour lines at
heights $c^2$ and $c^2+1$  of the first-order RWs of the complex mKdV and the NLS are plotted in Fig. 22. These pictures
and analytical formulae of length and width show that, for the first-order RWs of the complex mKdV and the NLS, they  are very similar to each other apart from a remarkable tilt with respect to the axes and a remarkable shortening of length of them with same values of  $a$ and $c$.
In other words,  the inclusion of third-order dispersion and time-delay correction is responsible for a strong
rotation and a strong compression effects in the first-order RW of the complex mKdV equation. In particular, if $a=0$, $|q_{NLS}^{[1]}|2$ is still a
RW, but $|q^{[1]}|^2$ of the complex mKdV is a soliton traveling along $l_3$ which is no longer doubly-localized in $x$ and $t$ directions.

In terms of applications, the investigation of the above investigated
higher-order RW solutions will be useful to understand the
generation of high-power waves and their possible splitting,  etc. As
we have discussed earlier, very recently, using the non-slowly varying
envelope approximation (SVEA), the complex mKdV equation has been
derived and the generation of few-cycle optical pulses have been
reported\cite{leblond,leblond1}. In addition, it has been pointed
out that these type of few-cycle optical pulses require no phase
matching (a main issue in nonlinear optics), which makes a strong
contrast and provides an interesting aspect when compared with the longer
pulses derived by using the SVEA method. From these recent studies it is
also interesting to note that these type of few-cycle optical pulses
are very similar to the generation of high-power and very short
RW type ultra-short pulses. For example, in nonlinear
photonic crystal fibre, the above waves may be connected to the
generation of few-cycle optical pulses which will be useful to
realize the so-called supercontinuum generation. This type of white
light continuum coherence source will find a range of applications in
optical coherence tomography, optical meteorology, wavelength division
multiplexing, fluorescence microscopy, flow cytometry, atmospheric
sensing, etc.\cite{Agra,leblond,leblond1}.

\section{Conclusions}

In this paper, we applied the DT to construct the
higher-order RW-type rational solutions as well as the
evolution of rogue waves for the complex mKdV equation. Based on
detailed numerical and analytical investigations, we classified the
higher-order RWs with respect to their intrinsic structure.
We use the contour line method, for the first time to the best of our knowledge, to define the length and width of the first-order RW, and then provide their analytical formulae related to two parameters a and c. We illustrate clearly, by analytical formulas and figures, that the differences between the first-order RWs of the mKdV and the NLS are mainly due to strong rotation as well as strong compression effects. Furthermore, we observed that there are three
principle types, namely, fundamental pattern, ring pattern, and
triangular pattern. The composition of these three principle
patterns is mainly because of higher-order rogue waves. We also
provided several further new patterns of the higher-order RWs of
this model. The ring patterns obtained in this paper are similar to
the ``atom'' structure reported in \cite{akhmedievhigherorderrw2011}.
This explains the generalization and evolution of higher-order
RWs in terms of the solution. On the other hand, by changing the
free parameters in the DT, we have also constructed
more complicated (and interesting) structures. We deduced from our
stimulated examples in Figs. 16-20 that the non-standard decomposition
deserves further studies because there are presently unknown rules of the decomposition. Applying our
construction of RW solutions to different completely
integrable nonlinear evolution equations, it is interesting to
investigate some analogues between the evolution and decomposition
of higher-order RWs of these different integrable equations. It is
essential to find further conserved quantities for the
kind of solutions. These studies may help us for better
understanding of the occurrence of deep ocean waves with large
amplitude as well as the generation of few-cycle optical pulses
emitted  by high-power lasers which are used for the recently
invented supercontinuum generation sources, etc.

If we compare our results with the work in \cite{Zhaqilao} on
the rogue wave solutions of the complex mKdV, our results have
following advantages and developments:
\begin{itemize}
\item  Our method is considerably simpler as well as more systematic.
From Theorem 3, one can directly obtain the higher-order RWs without
calculating eigenfunctions $\psi^{[i]}_1$ and
$\phi^{[i]}_1(i=0,1,2,3)$ as in \cite{Zhaqilao}.
\item  We applied the contour line method to find the analytical description of the length and width of the first-order RW of the complex mKdV and the NLS equation.   We illustrated clearly, using suitable analytical
formulae and figures, that the differences between the first-order RWs of the mKdV and the NLS are due to a strong rotation and a strong compression effects. Note that, setting $a=0$,  $|q^{[1]}|^2$ reduces to a soliton on a background plane at height $c^2$, but  $|q_{NLS}^{[1]}|^2$ can not.

\item We  proposed and proved a convenient way to control the patterns and evolutions of the rogue wave by standard and non-standard decomposition with suitable choices of $s_i$.
\item We generated interesting patterns for 4th,5th and 6th rogue
waves.
\end{itemize}

With respect to the future research in this exciting area, we shall apply the contour line method to the first-order RW of the different NLS type equations. For the higher-order rogue waves, because the degree of polynomials in its explicit form is more than 4, it is not easy to get the analytical expressions of the asymptotes for its contour lines in general. Thus, how to get the analytical results on their length and width is an interesting, difficult and important problem, which deserves further study.


{\bf Acknowledgments}

{\noindent \small This work is supported by the NSF of China under
Grant No.11271210 and the K. C. Wong Magna Fund in Ningbo
University. J.S. He thanks sincerely Prof. A.S. Fokas for arranging
the visit to Cambridge University in 2012-2013 and for many useful
discussions. KP acknowledges  DST, NBHM, CSIR and IFCPAR, Government of India,
for the financial support through major projects. RE acknowledges
M. K\'eray for patient encouragement and is also grateful to NSF,
Hungary (OTKA, Ref. No.K83133).  This work has been partially
supported by The University of Sheffield's MSRC Visitor Grant.
We thank editorial board member for his/her suggestions on
our submission which has improved the clarity of the paper.}



\begin{table}[!ht]\label{distribution}
\caption{{\small Distribution of peaks on rings by non-standard
decomposition }  }
\begin{center}
\begin{tabular}{|c|c|c|c|c|}
  \hline
\bfseries Order & \multicolumn{4}{|c|}{\rule[-0.3cm]{0mm}{0.8cm}{\bfseries  Distributions of peaks on rings}
\hspace{1cm}(L=Left,R=Right)} \\
\hline
  4 &5+5({\tiny Fig.12}) &  &  & \\
  \hline
  5 & 7+7+1({\tiny Fig.13L}) &  5+5+5({\tiny Fig.13R})&  & \\
  \hline
  6 & 9+9+3({\tiny Fig.14R}) & 5+10+5+1({\tiny Fig.15L}) & 7+7+7({\tiny Fig.15R}) & 11+5+5({\tiny Fig.16})\\
  \hline
\end{tabular}
\end{center}
\end{table}

{\bf Appendix A:} In eq.(\ref{2ndrw}),  $L_1$ are $L_2$ are given by
\begin{eqnarray*}
B:=\mbox{\hspace{-0.7cm}}
&&2239488a^4c^{14}t^6+46656a^{12}c^6t^6+2985984c^{18}t^6+559872a^8c^{10}t^6\\
&&+746496a^6c^{10}t^5x+93312a^{10}c^6t^5x-1492992a^4c^{12}t^5x-186624a^8c^8t^5x\\
&&+1492992a^2c^{14}t^5x-2985984c^{16}t^5x+1244160c^{14}t^4x^2-248832a^6c^8t^4x^2\\
&&-995328a^2c^{12}t^4x^2+622080a^4c^{10}t^4x^2+77760a^8c^6t^4x^2-276480c^{12}t^3x^3\\
&&+34560a^6c^6t^3x^3-124416a^4c^8t^3x^3+248832a^2c^{10}t^3x^3+34560c^{10}t^2x^4\\
&&+8640a^4c^6t^2x^4-27648a^2c^8t^2x^4+1152a^2c^6tx^5-2304c^8tx^5+64c^6x^6\\
&&-331776c^6a^6t^4-518400c^{12}t^4+155520c^8a^4t^4-11664c^4a^8t^4+290304c^{10}t^3x\\
&&-15552c^4a^6t^3x+103680c^8a^2t^3x-176256c^6a^4t^3x-58752c^8t^2x^2-6912c^6a^2t^2x^2\\
&&-7776c^4a^4t^2x^2+4992c^6tx^3-1728c^4a^2tx^3-144c^4x^4+124416ac^{10}s_{1}t^3\\
&&+31104a^5c^6s_{1}t^3-165888a^3c^8s_{1}t^3+20736a^3c^6s_{1}t^2x-41472ac^8s_{1}t^2x\\
&&+3456ac^6s_{1}tx^2-18000c^6t^2-1620c^2a^4t^2-1080c^2a^2tx+5616c^4tx-180c^2x^2\\
&&+864c^4as_{1}t+144c^4s_{1}^2+45+i(2985984ac^{14}t^5+1492992a^5c^{10}t^5\\
&&+186624a^9c^6t^5-497664a^5c^8t^4x+995328a^3c^{10}t^4x-1990656ac^{12}t^4x\\
&&+248832a^7c^6t^4x-331776a^3c^8t^3x^2+124416a^5c^6t^3x^2+497664ac^{10}t^3x^2\\
&&-55296ac^8t^2x^3+27648a^3c^6t^2x^3+2304ac^6tx^4+207360c^8at^3-31104c^4a^5t^3\\
&&-13824c^6at^2x-20736c^4a^3t^2x-3456c^4atx^2+20736c^8s_{1}t^2-41472a^2c^6s_{1}t^2\\
&&+5184a^4c^4s_{1}t^2+3456a^2c^4s_{1}tx-6912c^6s_{1}tx+576c^4s_{1}x^2-2160ac^2t+144s_{1}c^2)\\
\end{eqnarray*}
and
\begin{eqnarray*}
C:=\mbox{\hspace{-0.7cm}}
&&2239488a^4c^{14}t^6+2985984c^{18}t^6+46656a^{12}c^6t^6+559872a^8c^{10}t^6\\
&&+746496a^6c^{10}t^5x-1492992a^4c^{12}t^5x-186624a^8c^8t^5x-2985984c^{16}t^5x\\
&&+93312a^{10}c^6t^5x+1492992a^2c^{14}t^5x+1244160c^{14}t^4x^2-995328a^2c^{12}t^4x^2\\
&&+622080a^4c^{10}t^4x^2+77760a^8c^6t^4x^2-248832a^6c^8t^4x^2+248832a^2c^{10}t^3x^3\\
&&+34560a^6c^6t^3x^3-124416a^4c^8t^3x^3-276480c^{12}t^3x^3-27648a^2c^8t^2x^4\\
&&+8640a^4c^6t^2x^4+34560c^{10}t^2x^4-2304c^8tx^5+1152a^2c^6tx^5+64c^6x^6\\
&&+995328c^{10}a^2t^4+279936c^8a^4t^4-82944c^6a^6t^4+3888c^4a^8t^4-269568c^{12}t^4\\
&&+124416c^{10}t^3x+5184c^4a^6t^3x-51840c^6a^4t^3x-145152c^8a^2t^3x\\
&&-17280c^8t^2x^2-6912c^6a^2t^2x^2+2592c^4a^4t^2x^2+384c^6tx^3\\
&&+576c^4a^2tx^3+48c^4x^4-165888a^3c^8s_{1}t^3+124416ac^{10}s_1t^3\\
&&+31104a^5c^6s_{1}t^3+20736a^3c^6s_{1}t^2x-41472ac^8s_{1}t^2x+3456ac^6s_{1}tx^2\\
&&+20016c^6t^2+972c^2a^4t^2+6912c^4a^2t^2+648c^2a^2tx-2448c^4tx+108c^2x^2\\
&&-2592c^4as_{1}t+144c^4s_{1}^2+9.
\end{eqnarray*}

{\bf Appendix B:} In eq.(\ref{3rd}),  $L_1$ are $L_2$ are given by
\begin{eqnarray*}
L_1:=\mbox{\hspace{-0.7cm}}
&&-4939273445868140625t^{12}-545023276785450000t^{11}x-388407392651700000t^{10}x^2\\
&&-34025850934560000t^9x^3-12374529519456000t^8x^4-841946352721920t^7x^5\\
&&-204871837925376t^6x^6-10322713903104t^5x^7-1860148592640t^4x^8-62710087680t^3x^9\\
&&-8776581120t^2x^{10}-150994944tx^{11}-16777216x^{12}+2300237292280725000t^{10}\\
&&+500080291038360000t^9x+178207653254544000t^8x^2+20160748631347200t^7x^3\\
&&+4231975119851520t^6x^4+253682249269248t^5x^5+40317552230400t^4x^6+880347709440t^3x^7\\
&&+141203865600t^2x^8-1447034880tx^9+75497472x^{10}+257120426548112400t^8\\
&&-88521031030049280t^7x-1841385225323520t^6x^2-617799343104000t^5x^3\\
&&-88747774156800t^4x^4-8252622766080t^3x^5-200693514240t^2x^6+1415577600tx^7\\
&&+235929600x^8+12647412412496640t^6+42148769126400t^5x-2996161228800t^4x^2\\
&&-15902996889600t^3x^3 +313860096000t^2x^4-8139571200tx^5+707788800 x^6\\
&&-149676507590400t^4+2148738969600t^3x-1622998425600t^2x^2+31186944000t^3x\\
&&-928972800x^4-95215564800t^2+21598617600tx-464486400x^2+58060800\\
&&+i(-6540279321425400000t^{11}-601404995073600000t^{10}x-423057306879360000t^9x^2\\
&&-29901007761408000t^8x^3-10648770814771200t^7x^4-552411244265472t^6x^5\\
&&-130559642173440t^5x^6-4494741995520t^4x^7-779700142080t^3x^8-13589544960t^2x^9\\
&&-1811939328tx^{10}-303419904173280000t^9+263517097430016000t^8x\\
&&+29562711599923200t^7x^2+7820253397647360t^6x^3+799710056939520t^5x^4\\
&&+58500443013120t^4x^5+5243865661440t^3x^6+40768634880t^2x^7+6794772480tx^8\\
&&+10822374648023040t^7-15805156998758400t^6x-366298181959680t^5x^2\\
&&+157459297075200t^4x^3-6736379904000t^3x^4-249707888640t^2x^5+16986931200tx^6\\
&&+2321279745269760t^5-164322282700800t^4x+7849554739200t^3x^2-700710912000t^2x^3\\
&&+38220595200tx^4+4992863846400t^3-967458816000t^2x-33443020800tx^2-8360755200t)\\
L_2:=\mbox{\hspace{-0.7cm}}
&&4939273445868140625t^{12}+545023276785450000t^{11}x+388407392651700000t^{10}x^2\\
&&+34025850934560000t^9x^3+12374529519456000t^8x^4+841946352721920t^7x^5\\
&&+204871837925376t^6x^6+10322713903104t^5x^7+1860148592640t^4x^8\\
&&+62710087680t^3x^9+8776581120t^2x^{10}+150994944tx^{11}+16777216x^{12}\\
&&+1671541529351175000t^{10}-201221180459640000t^9x+29583374214960000t^8x^2\\
&&-8646206984294400t^7x^3-205872145551360t^6x^4-102185078390784t^5x^5\\
&&-5361951375360t^4x^6-141416202240t^3x^7-16349921280t^2x^8+2202009600tx^9\\
&&+25165824x^{10}+214192109547903600t^8+4346367735790080t^7x\\
&&+3783154346910720t^6x^2-1075635551969280t^5x^3+40942170316800t^4x^4\\
&&+1840480911360t^3x^5+84333035520t^2x^6+849346560tx^7+141557760x^8\\
&&+3537485306138880t^6+789853361080320t^5x+130057245388800t^4x^2\\
&&-11222478028800t^3x^3+402245222400t^2x^4-4034396160tx^5+613416960x^6\\
&&+90779142700800t^4-6216180019200t^3x-269186457600t^2x^2+11988172800tx^3\\
&&+221184000x^4+213178521600t^2-5009817600tx+199065600x^2+8294400.\\
\end{eqnarray*}
\clearpage

\begin{figure}[!ht]
\centering
\renewcommand{\figurename}{Fig.}
\includegraphics[height=7cm,width=8cm]{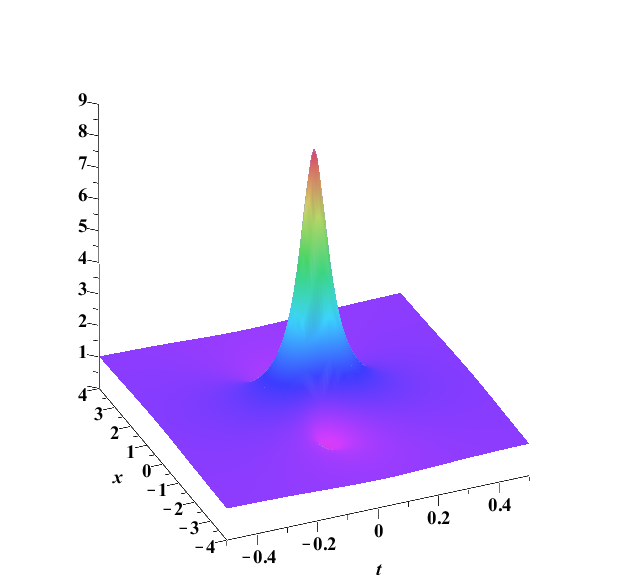}%
\hspace{1cm}
 \includegraphics[height=6cm,width=6cm]{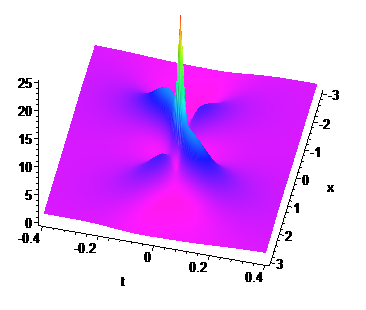}%
 \begin{center}
\parbox{15.5cm}{\small{\bf Figure 1.} (Color online)\ \
Fundamental pattern of the rogue wave.  (a) The left figure is the
evolution of $|q^{[1]}|^2$ ({\bf 1st-order} rogue wave)
with specific parameters $a=1.5,c=1,s_0=0$. (b) The right figure is
$|q^{[2]}|^2$ ({\bf 2nd-order} rogue wave) with specific parameters
$a=1.44,c=1,s_0=0,s_1=0$. }
\end{center}
\end{figure}
\begin{figure}[!ht]
\centering
\renewcommand{\figurename}{Fig.}
\includegraphics[height=6cm,width=6cm]{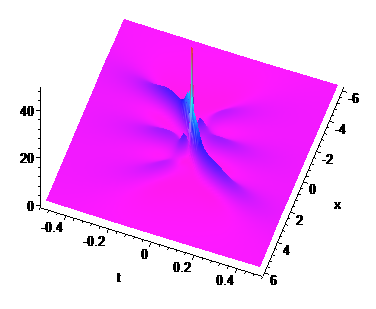}%
\hspace{1cm}
 \includegraphics[height=6cm,width=6cm]{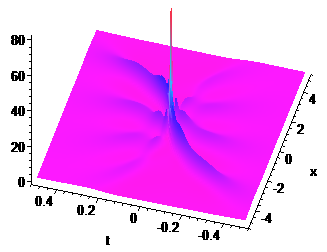}%
 \begin{center}
\parbox{15.5cm}{\small{\bf Figure 2.} (Color online)\ \ Fundamental pattern of the rogue
 wave. (a) The left figure is the evolution of $|q^{[3]}|^2$ ({\bf
3rd-order} rogue wave) with specific parameters
$a=1.4,c=1,s_0=0,s_1=0,s_2=0$. (b) The right figure is
$|q^{[4]}|^2$ ({\bf 4th-order} rogue wave) with specific parameters
$a=1.48,c=1,s_0=0,s_1=0,s_2=0,s_3=0$.}
\end{center}
\end{figure}
\begin{figure}[!ht]
\centering
\renewcommand{\figurename}{Fig.}
\includegraphics[height=6cm,width=6cm]{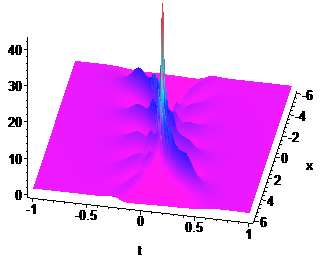}%
\hspace{1cm}
 \includegraphics[height=6cm,width=6cm]{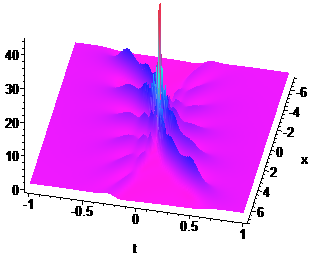}%
 \begin{center}
\parbox{15.5cm}{\small{\bf Figure 3.} (Color online)\ \ Fundamental patterns of the rogue waves.
(a) The left figure is the evolution of $|q^{[5]}|^2$ ({\bf
5th-order} rogue wave) with specific parameters
$a=1.5,c=1,s_0=0,s_1=0,s_2=0,s_3=0,s_4=0$. (b) The right figure is
$|q^{[6]}|^2$ ({\bf 6th-order} rogue wave) with specific parameters
$a=1.5,c=1,s_0=0,s_1=0,s_2=0,s_3=0,s_4=0,s_5=0$.}
\end{center}
\end{figure}

\begin{figure}[!ht]
\centering
\renewcommand{\figurename}{Fig.}
\includegraphics[height=7cm,width=7cm]{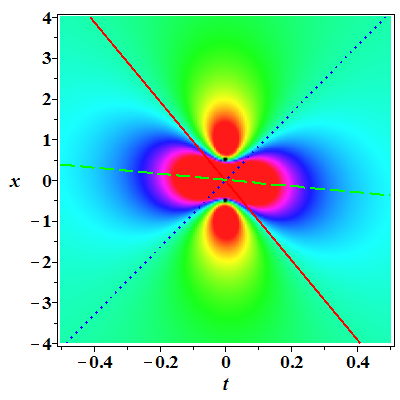}%
\hspace{0.1cm}
 \includegraphics[height=7cm,width=7cm]{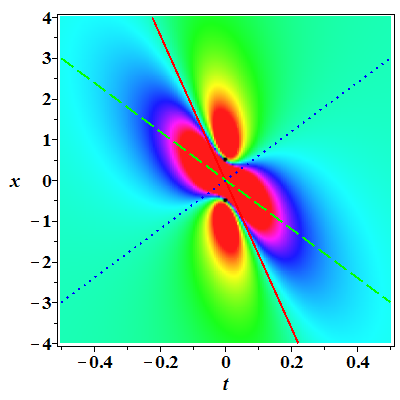}%
 \hspace{0.1cm}
 \includegraphics[height=7cm,width=7cm]{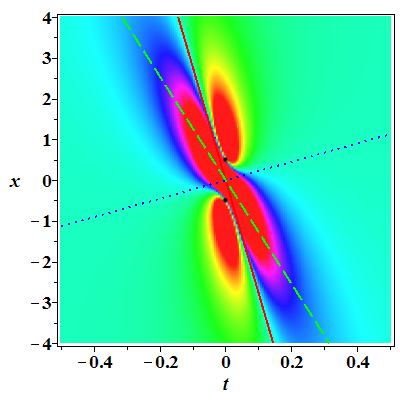}
 \begin{center}
\parbox{15.5cm}{\small{\bf Figure 4.} (Color online)\ \
The density plots of the first-order rogue wave $|q^{[1]}|^2$  with  $c=1$ and  different values of $a$.
From left to right, $a=1.5,2, 2.5$ in order.  Here red (solid) and blue (dot) lines are plotted for two asymptotes of contour lines at height $c^2$, green (dash) line is plotted for a median of one triangle composed of above two asymptotes and a parallel line
 of $x$-axis except $t=0$. Two fixed pints are located at $(0,0.5)$ and $(0,-0.5)$ in three panels.}
\end{center}
\end{figure}

\begin{figure}[!ht]
\centering
\renewcommand{\figurename}{Fig.}
\includegraphics[height=7cm,width=7cm]{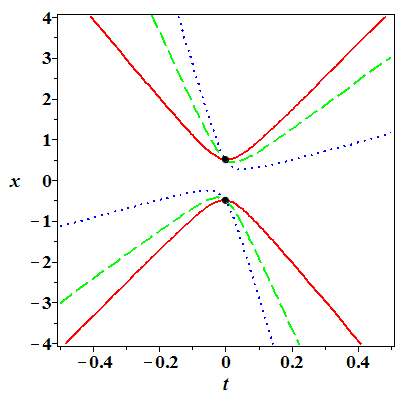}%
\hspace{0.2cm}
 \includegraphics[height=7cm,width=7cm]{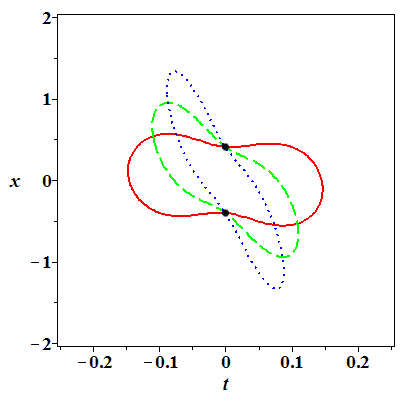}%
 \hspace{0.2cm}
 \includegraphics[height=7cm,width=7cm]{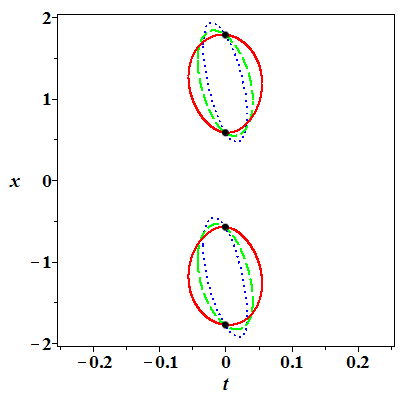}%
 \begin{center}
\parbox{15.5cm}{\small{\bf Figure 5.} (Color online)\ \
Contour lines of the first-order rogue wave with $c=1$ and $a=1.5$(red, solid), $2$(dash,green), $2.5$(dot, blue).
From left to right, panels are plotted for contour lines of  $|q^{[1]}|^2$ at height $c^2$ (on asymptotic plane),
 $c^2+1$, $c^2/2$ in order. There  are fixed points located at $(0,0.50)$ and $(0,-0.50)$ in the left panel,  $(0,0.41)$ and $(0,-0.41)$ in the middle panel,  $(0,1.78)$ and $(0,0.58)$, $(0,-0.58)$ and $(0,-1.78)$ in the right panel.}
\end{center}
\end{figure}

\begin{figure}[!ht]
\centering
\renewcommand{\figurename}{Fig.}
\includegraphics[height=7cm,width=7cm]{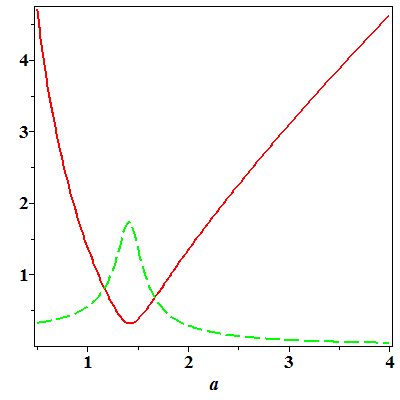}%
 \begin{center}
\parbox{15.5cm}{\small{\bf Figure 6.} (Color online)\ \
The length (red, solid) and the width (dash, green) of the first-order RW $|q^{[1]}|^2$ with fixed $c=1$. Note $a=\sqrt{2}$ is an extreme point of $d_L$ and $d_W$.}
\end{center}
\end{figure}

\begin{figure}[!ht]
\centering
\renewcommand{\figurename}{Fig.}
 \includegraphics[height=6cm,width=6cm]{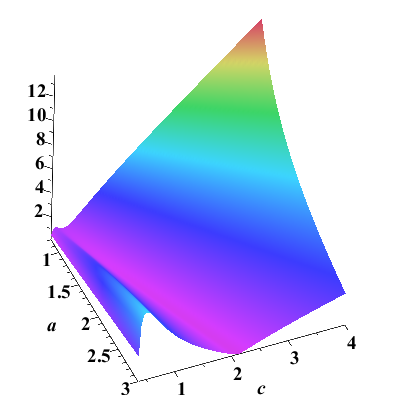}%
\hspace{1cm}
 \includegraphics[height=6cm,width=6cm]{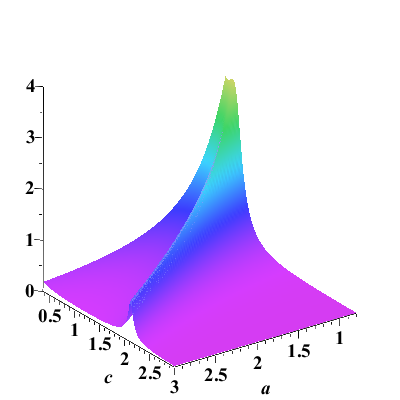}
 \begin{center}
\parbox{15.5cm}{\small{\bf Figure 7.} (Color online)\ \ The length(left) and width(right) of the first-order RW $|q^{[1]}|^2$  with
two parameters $a$ and $c$. }
\end{center}
\end{figure}

\begin{figure}[!ht]
\centering
\renewcommand{\figurename}{Fig.}
 \includegraphics[height=6cm,width=6cm]{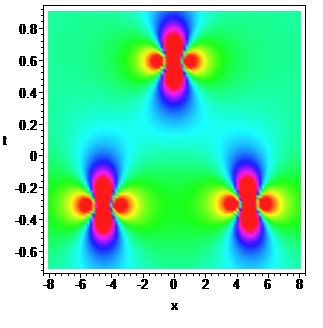}%
\hspace{1cm}
 \includegraphics[height=6cm,width=6cm]{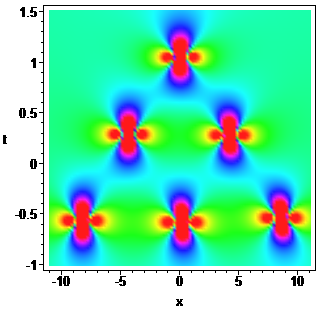}
 \begin{center}
\parbox{15.5cm}{\small{\bf Figure 8.} (Color online)\ \ Triangular pattern of the
2nd- and 3rd-order rogue waves.  The left panel is a
 density plot of  $|q^{[2]}|^2$ ({\bf 2nd-order} rogue wave) with $a=1.44,c=1,s_0=0,s_1=100$.  The right panel is a
 density plot of  $|q^{[3]}|^2$ ({\bf 3rd-order} rogue wave) with
 $a=1.44,c=1,s_0=0,s_1=100,s_2=0$. }
\end{center}
\end{figure}

\begin{figure}[!ht]
\centering
\renewcommand{\figurename}{Fig.}
\hspace{1cm}
 \includegraphics[height=6cm,width=6cm]{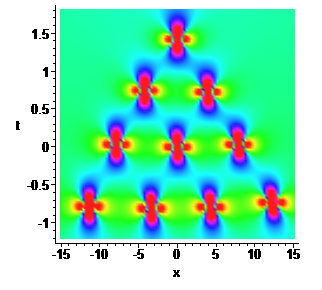}%
\hspace{1cm}
\includegraphics[height=6cm,width=6cm]{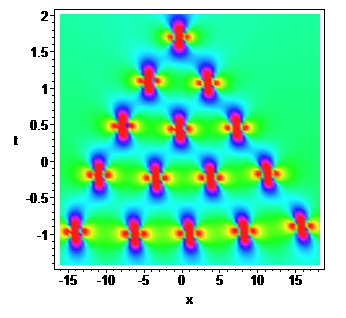}
 \begin{center}
\parbox{15.5cm}{\small{\bf Figure 9.} (Color online)\ \ Triangular pattern of the
4th-order and 5th-order rogue waves. The left panel is a
 density plot of  $|q^{[4]}|^2$ ({\bf 4th-order} rogue wave) with
  $a=1.46,c=1,s_0=0,s_1=100,s_2=0,s_3=0$, the right panel is
a density plot of  $|q^{[5]}|^2$ ({\bf 5th-order} rogue wave) with
$a=1.5,c=1,s_0=0,s_1=100,s_2=0,s_3=0,s_4=0 $.}
\end{center}
\end{figure}

\begin{figure}[!ht]
\centering
\renewcommand{\figurename}{Fig.}
\includegraphics[height=6cm,width=6cm]{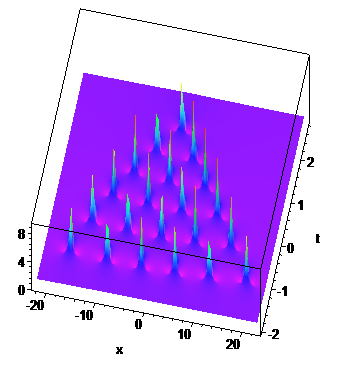}%
\hspace{1cm}
 \includegraphics[height=6cm,width=6cm]{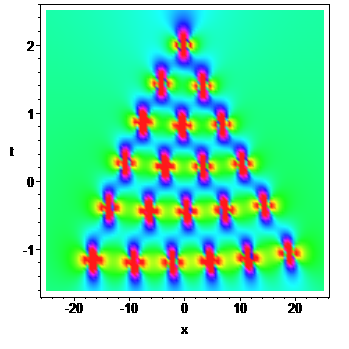}%
 \begin{center}
\parbox{15.5cm}{\small{\bf Figure 10.} (Color online)\ \ Triangular pattern of the
6th-order rogue wave. (a) The left figure is the evolution
of $|q^{[6]}|^2$ ({\bf 6th-order} rogue wave) and (b)the right figure
is the corresponding density plot with
$a=1.5,c=1,s_0=0,s_1=100,s_2=0,s_3=0,s_4=0,s_5=0$. }
\end{center}
\end{figure}
\newpage
\begin{figure}[!ht]
\centering
\renewcommand{\figurename}{Fig.}
 \includegraphics[height=6cm,width=6cm]{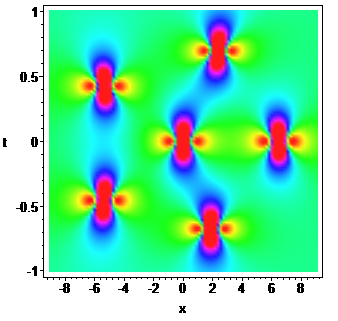}%
\hspace{1cm}
 \includegraphics[height=6cm,width=6cm]{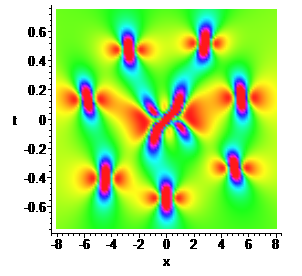}
 \begin{center}
\parbox{15.5cm}{\small{\bf Figure 11.} (Color online)\ \
Standard circular decomposition of the rogue wave: inner peak
surrounded by ring pattern.  The left panel is a
 density plot of  $|q^{[3]}|^2$ ({\bf 3rd-order} rogue wave) with
 $a=1.4,c=1,s_0=0,s_1=0,s_2=1000 $, the right panel is a
density plot of $|q^{[4]}|^2$ ({\bf 4th-order} rogue wave) with
$a=1.48,c=1,s_0=0,s_1=0,s_2=0,s_3=1000$. }
\end{center}
\end{figure}

\begin{figure}[!ht]
\centering
\renewcommand{\figurename}{Fig.}
 \includegraphics[height=6cm,width=6cm]{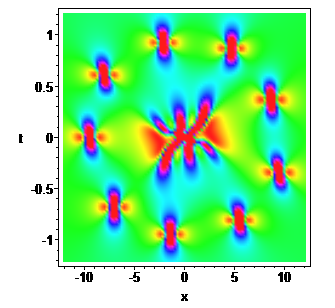}%
\hspace{1cm}
\includegraphics[height=6cm,width=6cm]{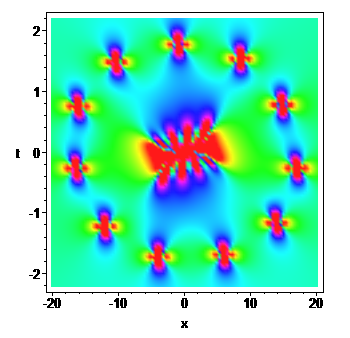}
 \begin{center}
\parbox{15.5cm}{\small{\bf Figure 12.} (Color online)\ \
Standard circular decomposition of the rogue wave: a ring pattern
with inner 3rd (left) and 4th (right) fundamental patterns. The left panel is a
 density plot of  $|q^{[5]}|^2$ ({\bf 5th-order} rogue wave) with
  $a=1.5,c=1,s_0=0,s_1=0,s_2=0,s_3=0,s_4=100000 $, the right
panel is a density plot of $|q^{[6]}|^2$ ({\bf 6th-order} rogue wave)
with $a=1.5,c=1,s_0=0,s_1=0,s_2=0,s_3=0,s_4=0,s_5=100000000 $.
}
\end{center}
\end{figure}
\newpage
\begin{figure}[!ht]
\centering
\renewcommand{\figurename}{Fig.}
 \includegraphics[height=6cm,width=6cm]{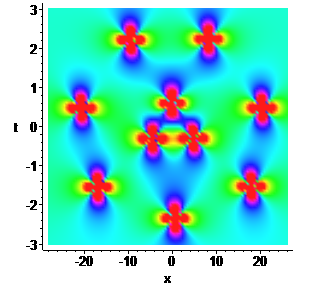}%
\hspace{1cm}
 \includegraphics[height=6cm,width=6cm]{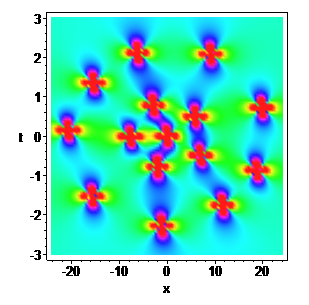}
 \begin{center}
\parbox{15.5cm}{\small{\bf Figure 13.} (Color online)\ \
Standard circular decomposition of the 4th and 5th rogue waves.
The left panel is a  density plot of  $|q^{[4]}|^2$ ({\bf 4th-order}
rogue wave) with $a=1.45,c=1,s_0=0,s_1=100,s_2=0,s_3=10000000 $ which is decomposed into
an outer ring with an inner 2nd triangular pattern,  the right
panel is a density plot of $|q^{[5]}|^2$({\bf 5th-order} rogue wave)
with $a=1.5,c=1,s_0=0,s_1=100,s_2=0,s_3=500000,s_4=100000000$
which is decomposed into an inner peak surrounded by two rings. }
\end{center}
\end{figure}
\begin{figure}[!ht]
\centering
\renewcommand{\figurename}{Fig.}
 \includegraphics[height=6cm,width=6cm]{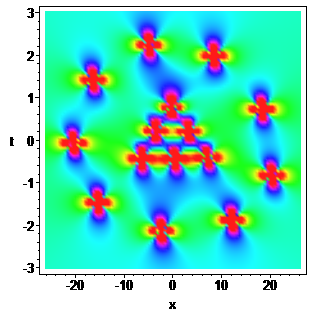}%
\hspace{1cm}
\includegraphics[height=6cm,width=6cm]{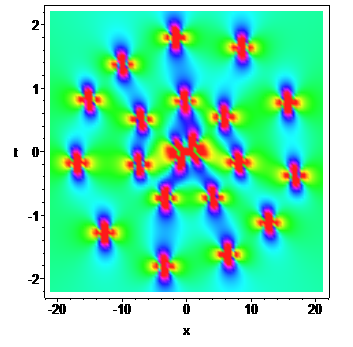}
 \begin{center}
\parbox{15.5cm}{\small{\bf Figure 14.} (Color online)\ \
Standard circular decomposition of the 5th and 6th rogue waves.
The left panel is a  density plot of  $|q^{[5]}|^2$ ({\bf 5th-order}
rogue wave) with $a=1.5,c=1,s_0=0,s_1=50,s_2=0,s_3=0,s_4=100000000 $ which is decomposed into
a ring pattern with an inner 3rd triangular pattern, the right
panel is a density plot of $|q^{[6]}|^2$ ({\bf 6th-order} rogue wave)
with $a=1.5,c=1,s_0=0,s_1=0,s_2=0,s_3=0,s_4=1000000,s_5=100000000$
which is decomposed into two ring patterns plus an inner 2nd fundamental pattern.}
\end{center}
\end{figure}

\begin{figure}[!ht]
\centering
\renewcommand{\figurename}{Fig.}
 \includegraphics[height=6cm,width=6cm]{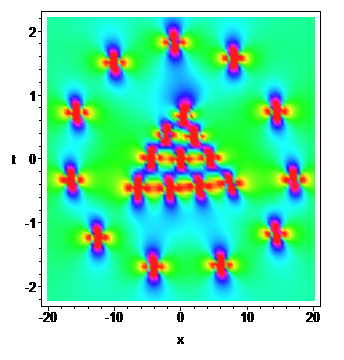}%
\hspace{1cm}
\includegraphics[height=6cm,width=6cm]{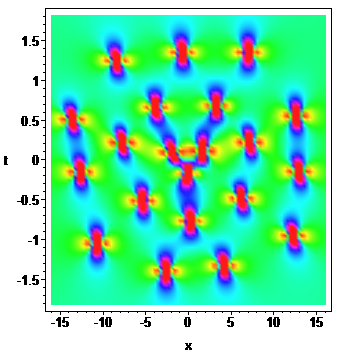}
 \begin{center}
\parbox{15.5cm}{\small{\bf Figure 15.} (Color online)\ \
Standard circular decomposition of the 6th rogue wave.
The left panel is a  density plot of  $|q^{[6]}|^2$ ({\bf 6th-order}
rogue wave) with $a=1.5,c=1,s_0=0,s_1=18,s_2=0,s_3=0,s_4=0,s_5=100000000 $ which is decomposed into
a ring plus an inner 4th triangular pattern, the right
panel is a density plot of $|q^{[6]}|^2$ ({\bf 6th-order} rogue wave)
with $a=1.5,c=1,s_0=0,s_1=0,s_2=0,s_3=7000,s_4=0,s_5=10000000 $
which is decomposed into three rings.}
\end{center}
\end{figure}
\begin{figure}[!ht]
\centering
\renewcommand{\figurename}{Fig.}
\includegraphics[height=6cm,width=6cm]{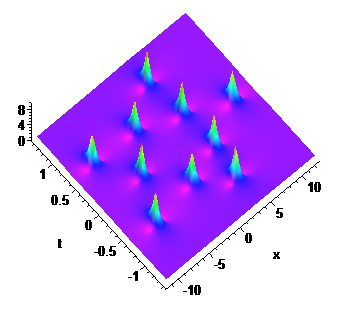}
\hspace{1cm}
 \includegraphics[height=6cm,width=6cm]{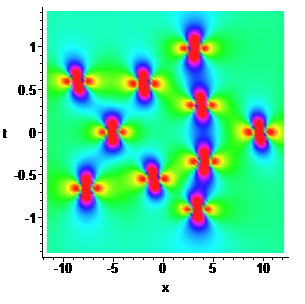}%
 \begin{center}
\parbox{15.5cm}{\small{\bf Figure 16.} (Color online)\ \
Non-standard circular decomposition of the 4th-rogue wave: two rings.
 (a) The left figure is the dynamical
evolution of $|q^{[4]}|^2$ ({\bf 4th-order} rogue wave) and (b) the
right figure is the corresponding density plot with
$a=1.48,c=1,s_0=0,s_1=0,s_2=1000,s_3=0$.
}
\end{center}
\end{figure}
\clearpage
\begin{figure}[!ht]
\centering
\renewcommand{\figurename}{Fig.}
 \includegraphics[height=6cm,width=6cm]{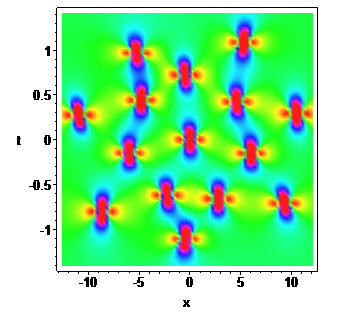}
\hspace{1cm}
 \includegraphics[height=6cm,width=6cm]{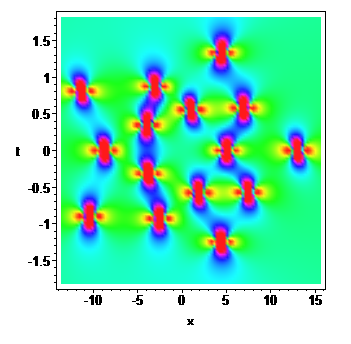}%
 \begin{center}
\parbox{15.5cm}{\small{\bf Figure 17.} (Color online)\ \
Non-standard circular decomposition of the 5th-rogue waves:
two rings plus an inner peak (left) and  three rings (right).
The left panel is a  density plot of  $|q^{[5]}|^2$ ({\bf 5th-order}
rogue wave) with $a=1.45,c=1,s_0=0,s_1=0,s_2=0,s_3=10000,s_4=0$,
the right
panel is a density plot of $|q^{[5]}|^2$ ({\bf 5th-order} rogue wave)
with $a=1.46,c=1,s_0=0,s_1=0,s_2=1000,s_3=0,s_4=0$. }
\end{center}
\end{figure}
\begin{figure}[!ht]
\centering
\renewcommand{\figurename}{Fig.}
 \includegraphics[height=6cm,width=6cm]{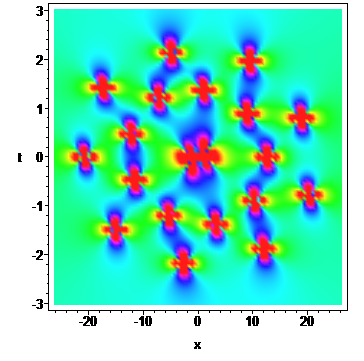}%
\hspace{1cm}
\includegraphics[height=6cm,width=6cm]{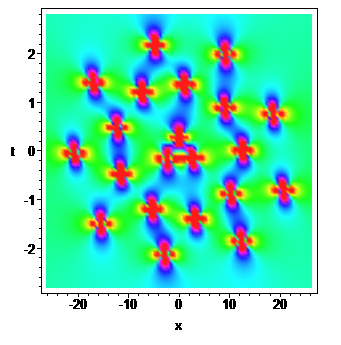}%
\begin{center}
\parbox{15.5cm}{\small{\bf Figure 18.} (Color online)\ \
Non-standard circular decomposition of the 6th-rogue wave: two ring
patterns with an inner 2nd fundamental pattern(left) and three rings(right).
The left panel is a  density plot of  $|q^{[6]}|^2$ ({\bf 6th-order}
rogue wave) with $a=1.5,c=1,s_0=0,s_1=0,s_2=0,s_3=0,s_4=10000000,s_5=0 $ , the right
panel is a density plot of $|q^{[6]}|^2$ ({\bf 6th-order} rogue wave)
with $a=1.5,c=1,s_0=0,s_1=15,s_2=0,s_3=0,s_4=10000000,s_5=0$.}
\end{center}
\end{figure}
\clearpage
\begin{figure}[!ht]
\centering
\renewcommand{\figurename}{Fig.}
 \includegraphics[height=6cm,width=6cm]{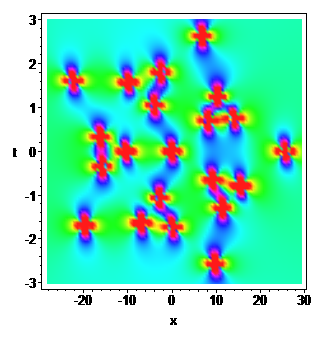}%
\hspace{1cm}
\includegraphics[height=6cm,width=6cm]{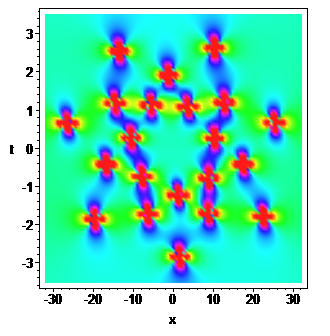}
 \begin{center}
\parbox{15.5cm}{\small{\bf Figure 19.} (Color online)\ \
Non-standard circular decomposition of the 6th-rogue wave: three rings
plus an inner peak (left) and  three rings (right).
The left panel is a  density plot of  $|q^{[6]}|^2$ ({\bf 6th-order}
rogue wave) with $a=1.5,c=1,s_0=0,s_1=0,s_2=10000,s_3=0,s_4=0,s_5=0 $ ,
 the right
panel is a density plot of $|q^{[6]}|^2$ ({\bf 6th-order} rogue wave)
with $a=1.5,c=1,s_0=0,s_1=0,s_2=0,s_3=1000000,s_4=0,s_5=100000000$. }
\end{center}
\end{figure}
\begin{figure}[!ht]
\centering
\renewcommand{\figurename}{Fig.}
\includegraphics[height=6cm,width=6cm]{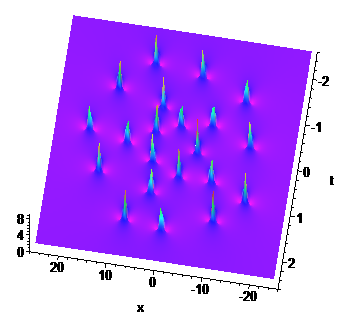}%
\hspace{1cm}
 \includegraphics[height=6cm,width=6cm]{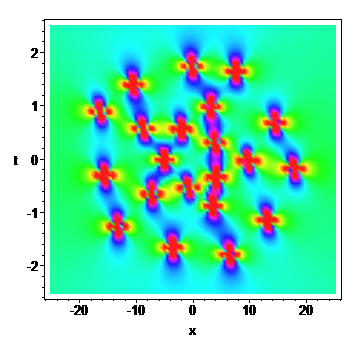}%
 \begin{center}
\parbox{15.5cm}{\small{\bf Figure 20.} (Color online)\ \
Non-standard circular decomposition of the 6th-rogue wave: three rings.
(a) The left panel is the dynamical
evolution of $|q^{[6]}|^2$ ({\bf 6th-order} rogue wave) and (b) the
right panel is the corresponding density plot with
$a=1.5,c=1,s_0=0,s_1=0,s_2=1000,s_3=1000000,s_4=0,s_5=100000000$. }
\end{center}
\end{figure}

\begin{figure}[!ht]
\centering
\renewcommand{\figurename}{Fig.}
\includegraphics[height=10cm,width=10cm]{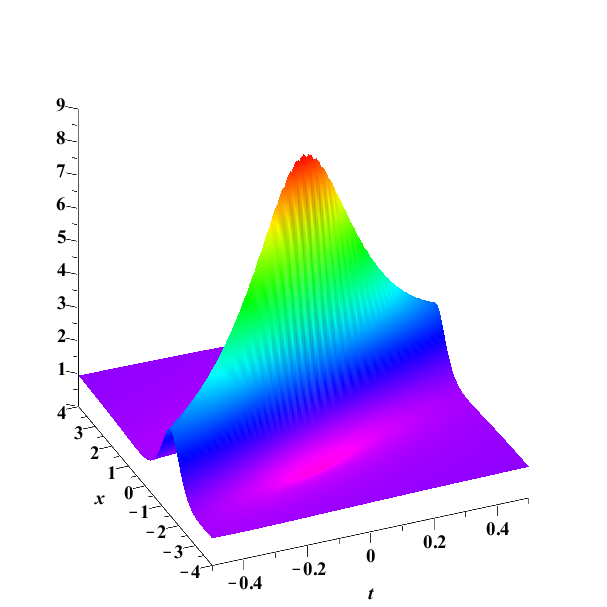}%
\hspace{1cm}
 \includegraphics[height=7cm,width=7cm]{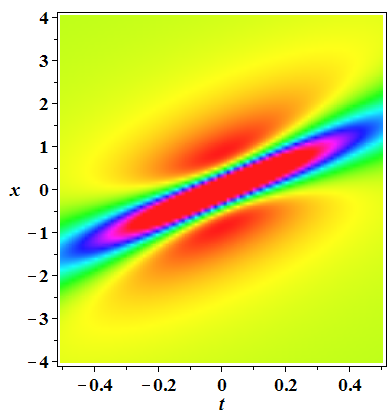}%
 \begin{center}
\parbox{15.5cm}{\small{\bf Figure 21.} (Color online)\ \
The first-order rogue  wave ($|q^{[1]}_{NLS}|^2$) of the NLS with $a=1.5$ and $c=1$.
The left panel is the dynamical evolution of $|q^{[1]}_{NLS}|^2$ and  the right panel is the corresponding density plot.
 }
\end{center}
\end{figure}
\begin{figure}[!ht]
\centering
\renewcommand{\figurename}{Fig.}
\includegraphics[height=6cm,width=6cm]{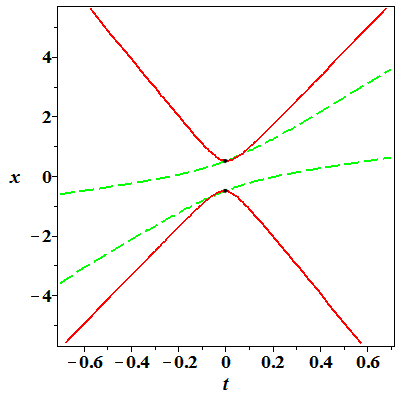}%
\hspace{1cm}
\includegraphics[height=6cm,width=6cm]{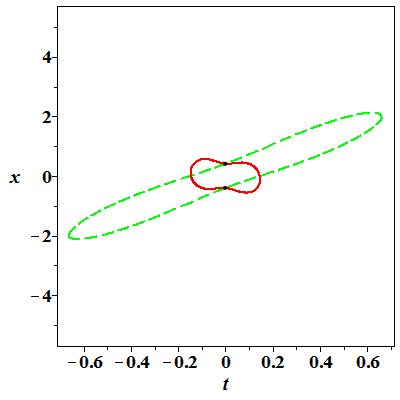}%
 \begin{center}
\parbox{15.5cm}{\small{\bf Figure 22.} (Color online)\ \ Contour lines of the first-order rogue wave  of the complex mKdV ($|q^{[1]}|^2$,red,solid) and  NLS  ($|q^{[1]}_{NLS}|^2$,green, dash) with $a=1.5$ and $c=1$. The left
panel is plotted at height $c^2$ (on the asymptotical plane), the right panel is plotted at height $c^2+1$.
 Two lines have two common pints:{$(0,0.50),(0,-0.50)$} in the left panel and {$(0,0.41),(0,-0.41)$} in the right panel.}
\end{center}
\end{figure}

\end{document}